\documentclass[11pt,onecolumn,aps,superscriptaddress, amsmath,amssymb,nofootinbib]{revtex4}   
\usepackage{latexsym}
\usepackage{graphics}

\usepackage{amsmath}    % need for subequations
\usepackage{graphicx}   % need for figures
\usepackage{verbatim}   % useful for program listings
\usepackage{color}      % use if color is used in text
\usepackage{subfigure}  % use for side-by-side figures
\usepackage{hyperref}   % use for hypertext links, including those to external documents and URLs
\newcommand{\bmat}{\left(\begin{array}}
\newcommand{\emat}{\end{array}\right)}
\newcommand{\beq}{\begin{equation}}
\newcommand{\eeq}{\end{equation}}
\newcommand{\drawsquare}[2]{\hbox{%
\rule{#2pt}{#1pt}\hskip-#2pt%  left vertical
\rule{#1pt}{#2pt}\hskip-#1pt%  lower horizontal
\rule[#1pt]{#1pt}{#2pt}}\rule[#1pt]{#2pt}{#2pt}\hskip-#2pt%  upper horizontal
\rule{#2pt}{#1pt}}% right vertical

\newcommand{\fund}{\raisebox{-.5pt}{\drawsquare{6.5}{0.4}}}%  fund
\newcommand{\Ysymm}{\raisebox{-.5pt}{\drawsquare{6.5}{0.4}}\hskip-0.4pt%
        \raisebox{-.5pt}{\drawsquare{6.5}{0.4}}}%  symmetric second rank
\newcommand{\Yasymm}{\raisebox{-3.5pt}{\drawsquare{6.5}{0.4}}\hskip-6.9pt%
        \raisebox{3pt}{\drawsquare{6.5}{0.4}}}%  antisymmetric second rank
\newcommand{\antifund}{\overline{\fund}}

\def\yzero{\smash{\hbox{$y\kern-4pt\raise1pt\hbox{${}^\circ$}$}}}

\def\-{\hphantom{-}}
\def\ov{\overline}
\def\s2{\frac{1}{\sqrt2}}

\def\beq{\begin{equation}}
\def\eeq{\end{equation}}
\def\beqa{\begin{eqnarray}}
\def\eeqa{\end{eqnarray}}

\def\IF{\relax{\rm I\kern-.18em F}}
\def\II{\relax{\rm I\kern-.18em I}}
\def\IP{\relax{\rm I\kern-.18em P}}

\def\Dsl{\,\raise.15ex\hbox{/}\mkern-13.5mu D} %can be subscripted

\def\IC{\bf C}
\def\IZ{\bf Z}
\def\IT{\bf T}
\def\z2z2{$\IC^3/(\IZ_2\times\IZ_2)$}

\def\s{\sigma}

\def\z{\zeta}

\def\bo{{\raise-.3ex\hbox{\large$\Box$}}}               % D'Alembertian
\def\face{{\raise.2ex\hbox{$\displaystyle \bigodot$}\mskip-2.2mu \llap {$\ddot
        \smile$}}}                                      % happy face
\def\leftrightarrowfill{$\mathsurround=0pt \mathord\leftarrow \mkern-6mu
        \cleaders\hbox{$\mkern-2mu \mathord- \mkern-2mu$}\hfill
        \mkern-6mu \mathord\rightarrow$}       % <--> double differential
\def\dvec#1{\vbox{\ialign{##\crcr
        \leftrightarrowfill\crcr\noalign{\kern-1pt\nointerlineskip}
        $\hfil\displaystyle{#1}\hfil$\crcr}}}           % <--> accent
\def\beq{\begin{equation}}
\def\eeq{\end{equation}}

\def\beqx{\begin{displaymath}}
\def\eeqx{\end{displaymath}}

\def\beqa{\begin{eqnarray}}
\def\eeqa{\end{eqnarray}}

\begin{document}

\title{
\normalsize \mbox{ }\hspace{\fill}
\begin{minipage}{12 cm}
%{\tt ~~~~~~~~~~~~~~~~~~
%UPR-1068-T,
 %hep-th/0403061}{\hfill}
\end{minipage}\\[5ex]
{\large\bf  Generalized Supersymmetric Pati-Salam Models \\
                 from Intersecting D6-branes
\\[1ex]}}
%\date{\today}

\author{Tianjun Li \footnote{E-mail: \texttt{tli@itp.ac.cn}}} 
\affiliation{CAS Key Laboratory of Theoretical Physics, Institute of Theoretical Physics,\\
	Chinese Academy of Sciences, Beijing 100190, P. R. China}
\affiliation{School of Physical Sciences, University of Chinese Academy of Sciences,\\
	No.19A Yuquan Road, Beijing 100049, P. R. China}
%\affiliation{
%	Institute of Theoretical Physics, Jiangxi Normal University, \\
%	Nanchang 330022, P.\ R.\ China\\
%}

\author{Adeel Mansha\footnote{E-mail: \texttt{adeelmansha@itp.ac.cn}}}
\affiliation{CAS Key Laboratory of Theoretical Physics, Institute of Theoretical Physics,\\
	Chinese Academy of Sciences, Beijing 100190, P. R. China}  
           \affiliation{School of Physical Sciences, University of Chinese Academy of Sciences,\\
	No.19A Yuquan Road, Beijing 100049, P. R. China}
  
\author{Rui Sun\footnote{E-mail: \texttt{sunrui@mpp.mpg.de}}}
\affiliation{Yau Mathematical Sciences Center, Tsinghua University, \\
	Haidian District, Beijing 100084, P. R. China}

% \thispagestyle{empty}

%\vspace*{-2cm}

\begin{abstract}
\medskip

Following the scanning methods of arXiv:1910.04530, 
we for the first time systematically construct the $N=1$ supersymmetric
 $SU(12)_C\times SU(2)_L\times SU(2)_R$ models, $SU(4)_C\times SU(6)_L\times SU(2)_R$ models, and $SU(4)_C\times SU(2)_L\times SU(6)_R$ models from the Type IIA orientifolds on $\IT^6/(\mathbb Z_2\times \mathbb Z_2)$ with intersecting D6-branes. These gauge symmetries can be broken down to the Pati-Salam gauge symmetry $SU(4)_C\times SU(2)_L \times SU(2)_R$ via three $SU(12)_C/SU(6)_L/SU(6)_R$ adjoint representation Higgs fields, and further down to the Standard Model (SM) via the D-brane splitting  and Higgs mechanism. Also, we obtain three families of the SM fermions,
and have the left-handed and right-handed three-family SM fermion unification in the $SU(12)_C\times SU(2)_L\times SU(2)_R$ models, the left-handed three-family SM fermion unification in the $SU(4)_C\times SU(6)_L\times SU(2)_R$ models, and the right-handed three-family SM fermion unification in the $SU(4)_C\times SU(2)_L\times SU(6)_R$ models. Moreover, the $SU(4)_C\times SU(6)_L\times SU(2)_R$ models and $SU(4)_C\times SU(2)_L\times SU(6)_R$ models are related by the  left and right gauge symmetry exchanging, as well as a variation of type II T-duality. Furthermore, the hidden sector contains $USp(n)$ branes, which are parallel with the orientifold planes or their $Z_2$ images and might break the supersymmetry via gaugino condensations.

\end{abstract} 
\maketitle

%\pacs{\tt PACS number(s): }
\vskip2cm
\newpage

\section{Introduction}

Constructing the $N=1$ supersymmetric Standard Models (SM) or SM from string theories has been the essential goal of string phenomenology. D-branes as boundaries of open strings plays an important role in phenomenologically interesting model building in Type I, Type IIA and Type IIB string theories~\cite{JPEW}. Conformal field theory provides the consistent constructions of four-dimensional supersymmetric $N=1$ chiral models with non-Abelian gauge symmetry on Type II orientifolds for the open string sectors. The chiral fermions on the worldvolume of the D-branes are located at orbifold singularities~\cite{ABPSS, berkooz, ShiuTye, lpt, MCJW, Ibanez, MKRR}, and/or at the intersections of D-branes in the internal space~\cite{bdl} with a T-dual description in terms of magnetized D-branes as shown in~\cite{bachas,urangac}.
Many non-supersymmetric three-family SM-like models and generalized unified models have been constructed [12$-$25], within the  intersecting D6-brane models on Type IIA orientifolds~\cite{bgkl, bkl, afiru}. These models typically suffer from the large Planck scale corrections at the loop level which results in the gauge hierarchy problem. A large number of the supersymmetric Standard-like models and generalized unified models have been constructed~\cite{CSU1, CSU2, CP, CPS, CLS1, CLS2, MCIP, CLW, blumrecent, Honecker,
LLG3, Cvetic:2004ui, Cvetic:2004nk, Chen:2005aba, Chen:2005mm, Chen:2005mj,
Chen:2007px, Chen:2007ms, Chen:2007zu, Chen:2008rx, Blumenhagen:2005mu}, with the above problem solved.
For a pedagogical introduction to phenomenologically interesting string models constructed with intersecting D-Branes, we refer to~\cite{Blumenhagen:2006ci}.

Along this direction, explicit models for the three-family $N=1$ supersymmetric Pati-Salam models with Type IIA orientifolds on $\IT^6/(\IZ_2\times \IZ_2)$ with intersecting D6-branes have been systematically constructed in ~\cite{Cvetic:2004ui}. The gauge symmetries all come from $U(n)$ branes, while
the Pati-Salam gauge symmetries $SU(4)_C\times SU(2)_L \times SU(2)_R$ 
break down to $SU(3)_C\times SU(2)_L \times U(1)_{B-L} \times U(1)_{I_{3R}} $ via D6-brane splittings. It further break down to the SM via four-dimensional $N=1$ supersymmetry via Higgs mechanism. This provides a way to realize the SM without any additional  anomaly-free $U(1)$'s around the electroweak scale introduced. Note that there are also hidden sector contain $USp(n)$ branes paralleling to the orientifold planes or their ${\bf Z_2}$ images. These models normally are constructed with at least two confining gauge groups in the hidden sector, for which the gaugino condensation triggers supersymmetry breaking and (some) moduli stabilization. In particular, one of these type of models is with a realistic phenomenology found by Chen, Mayes, Nanopoulos and one of us (TL) in~\cite{Chen:2007px, Chen:2007zu}. Its variations are also visited in~\cite{Chen:2007ms}. 
Moreover,  there are a few other potentially interesting constructions with possible massless 
vector-like  fields that might lead to SM~\cite{Cvetic:2004ui}. These vector fields are
 not arising from a $N=2$ subsector, but can break the Pati-Salam gauge symmetry down to the SM or break the $U(1)_{B-L}\times U(1)_{I_{3R}}$ down to $U(1)_Y$.  
 For such construction, large wrapping numbers are required because the increased absolute values of the intersection numbers between $U(4)_C$ stack of D-branes and $U(2)_R$ stack(or its orientifold image). Therefore,  more powerful scanning methods reaching large wrapping numbers are also requested in further investigations.

Employing our improved scanning methods and machine learning techniques, we systematically studied the three-family $N=1$ supersymmetric Pati-Salam model building in Type IIA orientifolds on $\IT^6/(\IZ_2\times \IZ_2)$ with intersecting D6-branes in which the $SU(4)_C\times SU(2)_L \times SU(2)_R$ gauge symmetries arise from $U(n)$ branes. In particular, we construct the new models with large winding numbers, and find that the approximate gauge coupling unification can be achieved at the string scale.
The Pati-Salam gauge symmetries $SU(4)_C\times SU(2)_L \times SU(2)_R$ therein can be broken down to the SM via D-brane splitting as well as D- and F-flatness preserving Higgs mechanism. The hidden sector contains $USp(n)$ branes with $n$ equals to 4 or 2, which are parallel with the orientifold planes or their ${\mathbb Z_2}$ images. We find that the Type II T-duality in the previous study~\cite{Cvetic:2004ui} is not an equivalent relation in Pati-Salam model building as most of the models are not invariant under $SU(2)_L$ and $SU(2)_R$ exchange. As follows, by swapping the b- and c-stacks of branes, gauge couplings can be redefined and refine the gauge unification becomes possible.
We systematically construct explicit new models with three families, which usually do not have gauge coupling unification at the string scale. We for the first time construct the Pati-Salam models with one large wrapping number reaching $5$.  In particular, we find that these models carry more refined gauge couplings, and with better approximate gauge coupling unification. With dimension reduction method ``LatentSemanticAnalysis'', we show that the three-family $N=1$ supersymmetric Pati-Salam models gather on islands, where more interesting models can be expected.

Following the scanning methods in Ref.~\cite{Li:2019nvi}, 
we for the first time systematically construct the $N=1$ supersymmetric
$SU(12)_C\times SU(2)_L\times SU(2)_R$ models, $SU(4)_C\times SU(6)_L\times SU(2)_R$ models, and $SU(4)_C\times SU(2)_L\times SU(6)_R$ models from the Type IIA orientifolds on $\IT^6/(\mathbb Z_2\times \mathbb Z_2)$ with intersecting D6-branes. These gauge symmetries can be broken down to the Pati-Salam gauge symmetry $SU(4)_C\times SU(2)_L \times SU(2)_R$ via three $SU(12)_C/SU(6)_L/SU(6)_R$ adjoint representation Higgs fields, and further down to the Standard Model (SM) via the D-brane splitting  and Higgs mechanism. Also, we obtain three families of the SM fermions,
and have the left-handed and right-handed three-family SM fermion unification in the $SU(12)_C\times SU(2)_L\times SU(2)_R$ models, the left-handed three-family SM fermion unification in the $SU(4)_C\times SU(6)_L\times SU(2)_R$ models, and the right-handed three-family SM fermion unification in the $SU(4)_C\times SU(2)_L\times SU(6)_R$ models. Moreover, the $SU(4)_C\times SU(6)_L\times SU(2)_R$ models and $SU(4)_C\times SU(2)_L\times SU(6)_R$ models are related by the  left and right gauge symmetry exchanging, as well as a variation of type II T-duality n Ref.~\cite{Li:2019nvi}, but the $U(1)_Y$ gauge coupling are different. Furthermore, the hidden sector contains $USp(n)$ branes, which are parallel with the orientifold planes or their $Z_2$ images and might break the supersymmetry via gaugino condensations.

This paper is organized as follows. We will firstly review the basic rules for supersymmetric intersecting D6-brane model building on Type IIA $T^6/(\IZ_2\times \IZ_2)$ orientifolds in Section II, as well as  the tadpole cancellation conditions and the conditions for D6-brane configurations which preserve four-dimensional $N=1$ supersymmetry in Section III. We present the generalized supersymmetric Pati-Salam model building in Section IV. We discuss the preliminary phenomenological consequences in Section V. The discussions and conclusion are in Section VI.

\section{$T^6 /(Z_2 \times Z_2)$ Orientifolds with Intersecting D6-Branes}
\label{sec:orientifold}

Before we construct the generalized verison of Pati-Salam models, let us briefly review the basic rules to construct the supersymmetric models  on Type IIA $T^6 /(Z_2 \times Z_2)$ orientifolds with D6-branes intersecting at generic angles to obtain the massless open string state spectra as in~\cite{CSU2, CPS}.
%In Type IIA string theory which is compactified on a $T^6 /(Z_2 \times Z_2)$ orientifold, 
In this construction,
we consider the six-torus $T^{6}$ factorized as three two-tori  $T^{6} = T^{2} \times T^{2} \times T^{2}$ with  complex coordinates for the $i$-th two-torus to be $z_i$, $i=1,\; 2,\; 3$ respectively.
The $\theta$ and $\omega$ generators for the orbifold group $Z_{2} \times Z_{2}$ are associated with the twist vectors $(1/2,-1/2,0)$ and $(0,1/2,-1/2)$ respectively. They act on the complex coordinates $z_i$ in the form of
\beqa
& \theta: & (z_1,z_2,z_3) \to (-z_1,-z_2,z_3)~,~ \nonumber \\
& \omega: & (z_1,z_2,z_3) \to (z_1,-z_2,-z_3)~.~\,
\label{orbifold} \eeqa 
Furthermore, we implement the orientifold projection by gauging the $\Omega R$ symmetry.
In which, $\Omega$ is world-sheet parity, and $R$ acts on the complex coordinates as
 \beqa
 R: (z_1,z_2,z_3) \to ({\ov z}_1,{\ov z}_2,{\ov
z}_3)~.~\, 
%\labelorientifold} 
\eeqa 
In total, there are four kinds of orientifold 6-planes (O6-planes) for the actions of $\Omega R$, $\Omega R\theta$, $\Omega R\omega$, and $\Omega R\theta\omega$ respectively. In addition,  three stacks of $N_a$ D6-branes wrapping on the factorized three-cycles are introduced to  cancel the RR charges of these O6-planes. 
As discussed in~\cite{bkl, Chen:2007zu, CSU2,CPS}, these two-tori are with two kinds of complex structures: rectangular and tilted, which are consistent with the orientifold projection. 
The homology classes of the three cycles which are wrapped by the D6-brane stacks takes the form $n_a^i[a_i]+m_a^i[b_i]$ and $n_a^i[a'_i]+m_a^i[b_i]$ for the rectangular and tilted tori respectively, with $[a_i']=[a_i]+\frac{1}{2}[b_i]$. Therefore, a generic one cycle are labelled by $(n_a^i,l_a^i)$ in terms of the wrapping numbers, $l_{a}^{i}\equiv m_{a}^{i}$ and $l_{a}^{i}\equiv 2\tilde{m}_{a}^{i}=2m_{a}^{i}+n_{a}^{i}$ for a rectangular and tilted two-torus, respectively. Moreover, $l_a^i-n_a^i$ is even for  tilted two-tori.

We note the wrapping number for stack $a$ of D6-branes along the cycle  to be $(n_a^i,l_a^i)$, and their $\Omega R$ images ${a'}$ stack of $N_a$ D6-branes are with wrapping numbers $(n_a^i,-l_a^i)$. 
The homology three-cycles 
for stack $a$ of D6-branes and  its orientifold image  $a'$  takes the form of 
\beq \label{homo}
[\Pi_a]=\prod_{i=1}^{3}\left(n_{a}^{i}[a_i]+2^{-\beta_i}l_{a}^{i}[b_i]\right),\;\;\;
\left[\Pi_{a'}\right]=\prod_{i=1}^{3}
\left(n_{a}^{i}[a_i]-2^{-\beta_i}l_{a}^{i}[b_i]\right)~,~\, \eeq
where $\beta_i=0$  for the rectangular and $\beta_i=1$ tilted $i$-th two-torus.
The homology three-cycles wrapped by the four O6-planes are in terms of
\beq \Omega R: [\Pi_{\Omega R}]= 2^3
[a_1]\times[a_2]\times[a_3]~,~\, \eeq \beq \Omega R\omega:
[\Pi_{\Omega
R\omega}]=-2^{3-\beta_2-\beta_3}[a_1]\times[b_2]\times[b_3]~,~\,
\eeq \beq \Omega R\theta\omega: [\Pi_{\Omega
R\theta\omega}]=-2^{3-\beta_1-\beta_3}[b_1]\times[a_2]\times[b_3]~,~\,
\eeq \beq
 \Omega R\theta:  [\Pi_{\Omega
R}]=-2^{3-\beta_1-\beta_2}[b_1]\times[b_2]\times[a_3]~.~\,
\label{orienticycles} \eeq 
The intersection numbers are related with  wrapping numbers in the term of
\beq \label{intersectionab}
I_{ab}=[\Pi_a][\Pi_b]=2^{-k}\prod_{i=1}^3(n_a^il_b^i-n_b^il_a^i)~,~\,
\eeq \beq  \label{intersectionab'}
I_{ab'}=[\Pi_a]\left[\Pi_{b'}\right]=-2^{-k}\prod_{i=1}^3(n_{a}^il_b^i+n_b^il_a^i)~,~\,
\eeq \beq  \label{intersectionaa'}
I_{aa'}=[\Pi_a]\left[\Pi_{a'}\right]=-2^{3-k}\prod_{i=1}^3(n_a^il_a^i)~,~\,
\eeq \beq {I_{aO6}=[\Pi_a][\Pi_{O6}]=2^{3-k}(-l_a^1l_a^2l_a^3
+l_a^1n_a^2n_a^3+n_a^1l_a^2n_a^3+n_a^1n_a^2l_a^3)}~,~\,
\label{intersections} \eeq 
where $k=\beta_1+\beta_2+\beta_3$ is the total number of the tilted two-tori, while $[\Pi_{O6}]=[\Pi_{\Omega R}]+[\Pi_{\Omega R\omega}]+[\Pi_{\Omega R\theta\omega}]+[\Pi_{\Omega R\theta}]$ is the sum of four O6-plane homology three-cycles.

On the model building side, the  massless particle spectrum for intersecting D6-branes at general angles can be expressed in terms of the intersection numbers shown in Table \ref{genspectrum}.
In this table, the representations refer to $U(N_a/2)$, the gauge symmetry results from $Z_2\times Z_2$ orbifold projection~\cite{CSU2}.  
The chiral supermultiplets contain both scalars and fermions in the supersymmetric constructions, while the positive intersection numbers refer to the left-handed chiral supermultiplets.
The two main constraints on the four-dimensional $N=1$ supersymmetric model building from Type IIA orientifolds with intersecting D6-branes are: RR tadpole cancellation conditions and 
$N=1$ supersymmetry preservation in four dimensions, which we will discuss in the following with our generalized construction.

\begin{table}[t]
	\label{genspectrum}
\caption{ 
General massless particle spectrum for intersecting D6-branes at generic angles.
 }
\renewcommand{\arraystretch}{1.25}
\begin{center}
\begin{tabular}{|c|c|}
\hline {\bf Sector} & \phantom{more space inside this box}{\bf
Representation}
\phantom{more space inside this box} \\
\hline\hline
$aa$   & $U(N_a/2)$ vector multiplet  \\
       & 3 adjoint chiral multiplets  \\
\hline
$ab+ba$   & $I_{ab}$ $(\fund_a,\antifund_b)$ fermions   \\
\hline
$ab'+b'a$ & $I_{ab'}$ $(\fund_a,\fund_b)$ fermions \\
\hline $aa'+a'a$ &$\frac 12 (I_{aa'} - \frac 12 I_{a,O6})\;\;
\Ysymm\;\;$ fermions \\
          & $\frac 12 (I_{aa'} + \frac 12 I_{a,O6}) \;\;
\Yasymm\;\;$ fermions \\
\hline
\end{tabular}
\end{center}
\label{spectrum}
\end{table}

\section{Generalized D6-Brane Constructions}

In the random scanning we performed for new model searching in~\cite{Li:2019nvi}, we observe that new models with three generations of particles can also be constructed when $n_x^i$ and $l_x^i$ are with common factor $3$, while $x$ refers to $a, b, c$ stacks of branes and $i$ refers to $1, 2, 3$ for different wrapping directions. In more details, for example when $n_a^1$ and $l_a^1$ are with common factor 3, the intersection numbers consequently will also have co-factor 3 for  Eqs.~\eqref{intersectionab},~\eqref{intersectionab'}, and co-factor 6 for Eq.~\eqref{intersectionaa'}. Divided by this co-factor 3 for the stack $a$ of D-brane, the generalized gauge group resulting from the D6-branes becomes gauge symmetries $SU(12)_C\times SU(2)_L \times SU(2)_R$. In which the a-stack brane's gauge $U(12)$ can be broken down to $U(4)$  with proper orientations. 
%Now we take $n_b^1, l_b^1$ are with co-factor 3 as example to discuss the gauge breaking to standard Pati-Salam gauge symmetries. 
When the wrapping number $n_b^1, l_b^1$ for the b-stack of brane are with common factor 3, by dividing out the co-factor 3,  the resulting gauge symmetry becomes $SU(4)_C\times SU(6)_L \times SU(2)_R$. To break it down to the Pati-Salam symmetry  $SU(4)_C\times SU(2)_L \times SU(2)_R$, similar orientations needed to be perfomed to break the $U(6)_L$ gauge to $U(2)_L$ which we will give details in section~\ref{sec:models}. 
Also, we obtain three families of the SM fermions,
and have the left-handed and right-handed three-family SM fermion unification in the $SU(12)_C\times SU(2)_L\times SU(2)_R$ models, the left-handed three-family SM fermion unification in the $SU(4)_C\times SU(6)_L\times SU(2)_R$ models, and the right-handed three-family SM fermion unification in the $SU(4)_C\times SU(2)_L\times SU(6)_R$ models.
As follows, the Pati-Salam gauge symmetries $SU(4)_C\times SU(2)_L \times SU(2)_R$  can be broken down to SM via D-brane splitting as well as D- and F-flatness preserving Higgs mechanism.	

As follows, we will present the two main constraints on the four-dimensional $N=1$ supersymmetric model building from Type IIA orientifolds with intersecting D6-branes, namely the RR tadpole cancellation conditions and $N=1$ supersymmetry preservation in four dimensions with our generalized gauge  modifications.

\subsection{The RR Tadpole Cancellation Conditions}

In the standard Pati-Salam models, the tadpole cancellation conditions lead to the $SU(N_a)^3$  cubic non-Abelian anomaly cancellation as shown in ~\cite{Uranga,imr,CSU2},  while the cancellation of $U(1)$ mixed gauge and gravitational anomaly\,(or $[SU(N_a)]^2 U(1)$ gauge anomaly) can be achieved by Green-Schwarz mechanism mediated by the untwisted RR fields as shown in~\cite{Uranga,imr,CSU2}.
The D6-branes and the orientifold O6-planes are the sources of RR fields and restricted by the Gauss law in a compact space. The sum of the RR charges from D6-branes  must cancel with it from the O6-planes due to the conservations of the RR field flux lines. The conditions 
for RR tadpole cancellations take the form of
\begin{eqnarray}  \label{tap1}
\sum_a N_a [\Pi_a]+\sum_a N_a
\left[\Pi_{a'}\right]-4[\Pi_{O6}]=0~,~\,
\end{eqnarray}
where the last term arises from the O6-planes are with $-4$ RR charges in D6-brane charge unit. 
To simplify the discussion of the following tadpole cancellation, we define the following products of wrapping numbers  as
\beq
\begin{array}{rrrr} 
A_a \equiv -n_a^1n_a^2n_a^3, & B_a \equiv n_a^1l_a^2l_a^3,
& C_a \equiv l_a^1n_a^2l_a^3, & D_a \equiv l_a^1l_a^2n_a^3, \\
\tilde{A}_a \equiv -l_a^1l_a^2l_a^3, & \tilde{B}_a \equiv
l_a^1n_a^2n_a^3, & \tilde{C}_a \equiv n_a^1l_a^2n_a^3, &
\tilde{D}_a \equiv n_a^1n_a^2l_a^3.\,
\end{array}
\label{variables}\eeq

In order to cancel the RR tadpoles,  an arbitrary number 
of D6-branes wrapping cycles along the orientifold planes need to be introduced. There are the so-called ``filler branes'', which contribute to the RR tadpole cancellation conditions, and trivially satisfy the four-dimensional $N=1$ supersymmetry conditions. In such way, the tadpole conditions can be written as
\begin{eqnarray}  \label{tap2}
 -2^k N^{(1)}+\sum_a N_a A_a=-2^k N^{(2)}+\sum_a N_a
B_a= \nonumber\\ -2^k N^{(3)}+\sum_a N_a C_a=-2^k N^{(4)}+\sum_a
N_a D_a=-16,\,
\end{eqnarray}
where $2 N^{(i)}$ is the number of filler branes wrapping along the $i$-th O6-plane. The filler branes representing the $USp$ group, carry the wrapping numbers as one of the O6-planes shown in Table \ref{orientifold}. The filler branes with non-zero $A$, $B$, $C$ or $D$ refer to  the $A$-, $B$-, $C$- or $D$-type $USp$ group, respectively. 

\renewcommand{\arraystretch}{1.4}
\begin{table}[t]
\caption{The wrapping numbers for four O6-planes.} \vspace{0.4cm}
\begin{center}
\begin{tabular}{|c|c|c|}
\hline
  Orientifold Action & O6-Plane & $(n^1,l^1)\times (n^2,l^2)\times
(n^3,l^3)$\\
\hline
    $\Omega R$& 1 & $(2^{\beta_1},0)\times (2^{\beta_2},0)\times
(2^{\beta_3},0)$ \\
\hline
    $\Omega R\omega$& 2& $(2^{\beta_1},0)\times (0,-2^{\beta_2})\times
(0,2^{\beta_3})$ \\
\hline
    $\Omega R\theta\omega$& 3 & $(0,-2^{\beta_1})\times
(2^{\beta_2},0)\times
(0,2^{\beta_3})$ \\
\hline
    $\Omega R\theta$& 4 & $(0,-2^{\beta_1})\times (0,2^{\beta_2})\times
    (2^{\beta_3},0)$ \\
\hline
\end{tabular}
\end{center}
\label{orientifold}
\end{table}

Note that for our generalized version of Pati-Salam models, the wrapping numbers $(n_x^i, l_x^i)$ wrap three times for one stack of D6-brane in one direction than the standard Pati-Salam model, the tadpole cancellation condition as shown in Eq.~\eqref{tap1}, \eqref{tap2} will also get rescaled in the relevant terms. It is obvious that when the first two terms in Eq.~\eqref{tap1} got rescaled with 3 factor, to satisfy the tadpole cancellation conditions, less number of orientifold planes are expected. This will also be confirmed with our random scanning method as constructed in our former study~\cite{Li:2019nvi}. We will present the generalized models in the next sections, and discuss its phenomenology aspects afterwards.

\subsection{Conditions for Four-Dimensional $N = 1$ Supersymmetric D6-Brane}

For the four-dimensional $N=1$ supersymmetric models,  the $1/4$ supercharges are required to be preserved from ten-dimensional Type I T-dual. Namely, under the orientation projection of the intersecting D6-branes and the $Z_2\times Z_2$ orbifold projection on the background manifold these $1/4$ supercharges are with the same value. 
It was shown in~\cite{bdl} that the four-dimensional $N=1$ supersymmetry can be preserved after the orientation projection iff the rotation angle of any D6-brane with respect to the orientifold plane is an element of $SU(3)$. Namely, $\theta_1+\theta_2+\theta_3=0 $ mod $2\pi$, where $\theta_i$ is the angle between the $D6$-brane and the orientifold-plane in the $i$-th two-torus. 
The 4-dimensional N = 1 supersymmetry will automatically survive the $Z_2\times Z_2$ orbifold projection~\cite{CPS}, and the SUSY conditions can therefore be written as 
\begin{eqnarray}
x_A\tilde{A}_a+x_B\tilde{B}_a+x_C\tilde{C}_a+x_D\tilde{D}_a=0,
\nonumber\\\nonumber \\ A_a/x_A+B_a/x_B+C_a/x_C+D_a/x_D<0,
\label{susyconditions}
\end{eqnarray} 
where $x_A=\lambda,\;
x_B=\lambda 2^{\beta_2+\beta3}/\chi_2\chi_3,\; x_C=\lambda
2^{\beta_1+\beta3}/\chi_1\chi_3,\; x_D=\lambda
2^{\beta_1+\beta2}/\chi_1\chi_2$, 
in which $\chi_i=R^2_i/R^1_i$ represent the complex structure moduli for the $i$-th two-torus.  
Moreover,  positive parameter $\lambda$  are introduced to put all the variables $A,\,B,\,C,\,D$ as equal footing. All the possible D6-brane configurations preserving four-dimensional $N=1$ supersymmetry can be classified into three types: 

%%%%%%%%%%%%%%%%%%%1058 1262

(1) Filler brane with the same wrapping numbers as one of the O6-planes from Table \ref{orientifold}. The gauge symmetry reveals to be $USp$ group. 
When there is only one of the wrapping number products $A$, $B$, $C$ or $D$ has non-zero and negative value, we refer to the $USp$ group as $A$-, $B$-, $C$- or $D$-type $USp$ group accordingly.

(2) When there are one zero wrapping number, two negative and two zero values 
in $A$, $B$, $C$ and $D$ we refer it as Z-type D6-brane.

(3) When there are three negative value and one positive value in $A$, $B$, $C$ and $D$ we refer to this case as NZ-type D6-brane. According to which one is positive, we note the NZ-type branes as $A$-, $B$-, $C$- and $D$-type NZ branes. Each type is with two forms of wrapping numbers are noted as follows
\begin{eqnarray}
A1: (-,-)\times(+,+)\times(+,+),& A2:(-,+)\times(-,+)\times(-,+);\\
B1: (+,-)\times(+,+)\times(+,+),& B2:(+,+)\times(-,+)\times(-,+);\\
C1: (+,+)\times(+,-)\times(+,+),& C2:(-,+)\times(+,+)\times(-,+);\\
D1: (+,+)\times(+,+)\times(+,-),& D2:(-,+)\times(-,+)\times(+,+).
\end{eqnarray}
%To be convenient, we shall refer the Z-type and NZ-type D6-branes to be $U$-branes in the following since they carry $U(n)$ gauge symmetry.

	For our generalized construction, the wrapping number $(n_x^i, l_x^i)$ will rescale the relevant terms in Eq.~\ref{susyconditions}, and thus the supersymmetry condition will need to be checked in the scaled manner. Moreover, although T-duality is not the focus of our work, we would like to note that 
	 if the three two-tori of two models and their corresponding wrapping numbers for all the D6-branes are correlated by an element of the permutation group $S_3$ acting on three two-tori, we call these	two models are equivalent. This applies to our generalized Pati-Salam Models in the same way. For more details about T-duality, D6-brane Sign Equivalent Principle and the equivalence of dual models, we refer to~\cite{Cvetic:2004ui, Li:2019nvi}.

\section{Generalized Supersymmetric Pati-Salam Model Building }
\label{sec:models}
%	and Gauge Symmetry Breaking via D6-Brane Splittings}
\subsection{Construction of Generalized Supersymmetric Pati-Salam Models}

In the standard Pati-Salam models, to construct the SM or SM-like models from the intersecting D6-brane scenarios, besides the $U(3)_C$ and $U(2)_L$ gauge symmetries from stacks of branes,  we construct two extra $U(1)$ gauge groups for both supersymmetric and non-supersymmetric models to have the correct quantum number for right-handed charged leptons as shown in~\cite{imr,CSU2,CPS,CP}. 
One  $U(1)_L$ represents the lepton number symmetry, while the other  $U(1)_{I_{3R}}$ behave as the third component of right-handed weak isospin. The hypercharge is then given by
\begin{eqnarray}
Q_Y=Q_{I_{3R}}+{{Q_B-Q_{L}}\over{2}}~,~\,
\end{eqnarray}
where $U(1)_B$ arises from the overall $U(1)$ in $U(3)_C$.
The $U(1)$ gauge symmetry\,(coming from a non-Abelian symmetry) is anomaly free and its gauge field is massless.
Thus, to forbid the gauge field of $U(1)_{I_{3R}}$ from obtaining a mass via $B\wedge F$ couplings,  $U(1)_{I_{3R}}$ can only arise from the non-Abelian part of $U(2)_R$ or $USp$ gauge symmetry. Similarly, to obtain an anomaly-free $U(1)_{B-L}$ gauge symmetry, the $U(1)_L$ gauge symmetry should come from non-Abelian group as well. 
Note that the $U(1)_L$ stack should be parallel to the $U(3)_C$ stack on at least one two-torus, we can obtain it by splitting one $U(4)$ stack of branes into $U(1)_L$ and $U(3)_C$
stacks.  The $U(3)_C$ gauge symmetry is also generated in the mean time.
When $U(1)_{I_{3R}}$ gauge arises from the stack of D6-branes on top of orientifold, there exist at least $8$ pairs of SM Higgs doublets, and two extra anomaly free $U(1)$ gauge symmetries  from the $USp$ group~\cite{CSU2,CP}. 
In which, these $U(1)$ gauge symmetries could be spontaneously broken by the Higgs mechanism via the scalar components of the chiral superfields with the quantum numbers of the right-handed neutrinos. 
However, they also break the D-flatness conditions and thus break supersymmetry. Therefore, the symmetry breaking scale is around the electroweak scale. We typically do not have any other candidates, which can preserve the D-flatness and F-flatness conditions, and break these gauge symmetries at an intermediate scale.

Distinguished from the model building in Ref.~\cite{Cvetic:2004ui, Li:2019nvi}, we for the first time construct a generalized version of Pati-Salam models with three times of wrapping for one stack of brane construction, then realize standard Pati-Salam models via Higgs mechanism. As follows, we can concentrate on the deriving Pati-Salam models in which $U(1)_{I_{3R}}$ arises from the $U(2)_R$ symmetry as usual.
As it is difficult to construct phenomenology interesting models with $SU(2)_L$ from the D6-branes on the top of O6-plane~\cite{Cvetic:2004ui}, 
we instead construct the gauge symmetries $SU(12)_C\times SU(2)_L\times SU(2)_R$, $SU(4)_C\times SU(6)_L\times SU(2)_R$, $SU(4)_C\times SU(2)_L\times SU(6)_R$ 
 from three time of wrapping respectively on a, b or c stack of D6-brane, which are not on the top of orientifold planes in a generalized construction. 
 %Because we do not have any extra anomaly free U(1) gauge symmetry around the electroweak scale, we solve a generic problem in previous constructions~\cite{CSU2,CP}.
Namely, we introduce three stacks of D6-branes, $a$, $b$, $c$ with D6-brane numbers 24, 4,  4;  8, 12, 4; and 8, 4, 12 which respectively give us the gauge symmetryies as above.
% $SU(12)_C\times SU(2)_L\times SU(2)_R$, $SU(4)_C\times SU(6)_L\times SU(2)_R$, and $SU(4)_C\times SU(2)_L\times SU(6)_R$. 
Then it can break down to three stacks of D6-branes, $a$, $b$, $c$ with D6-brane numbers $8, 4, 4$ with gauge symmetry $SU(4)_C\times SU(2)_L\times SU(2)_R$, and as follows break the resulting Pati-Salam gauge symmetry down to $SU(3)_C\times SU(2)_L\times U(1)_{B-L} \times U(1)_{I_{3R}}$ from D6-brane splitting. 
The SM gauge symmetry can be realized via Higgs mechanism with Higgs particles  from a $N=2$ subsector as for standard Pati-Salam models~\cite{Cvetic:2004ui}. 

Moreover, the gauge anomalies from three $U(1)$s are cancelled by the generalized Green-Schwarz mechanism, and these $U(1)$s gauge fields obtain masses via the linear $B\wedge F$ couplings.  Furthermore, to have three families of the SM fermions, we require the intersection numbers to satisfy
\begin{eqnarray}
\label{E3LF} I_{ab} + I_{ab'}~=~3~,~\,
\end{eqnarray}
\begin{equation}
\label{E3RF} I_{ac} ~=~-3~,~ I_{ac'} ~=~0~,~\,
\end{equation}
where the conditions $I_{ab} + I_{ab'}=3$ and $I_{ac} =-3$ give us three generations of the SM fermions,
whose quantum numbers under $SU(12)_C\times SU(2)_L\times SU(2)_R$  for example with gauge symmetries
are $({\bf 12, 2, 1})$ and $({\bf {\overline{12}}, 1, 2})$ in our generalized construction.  
Recall that the intersection numbers Eqs.~\eqref{intersectionab},~\eqref{intersectionab'},~\eqref{intersectionaa'} are in terms of the wrapping numbers, and our generalized construction are with three times of wrapping than standard construction of brane models, to satisfy the restriction of three families of SM fermions, it is expected that the three family models will be less but from higher energy level.
This also explains the generalized models with higher wrapping numbers are less populated in the model scanning according to wrapping numbers because of the common factor 3 in $(n_x^i, l_x^i)$. However,  common factor appears as 3, rather than other number, is natural to be understood because of the three family conditions are with factor 3 also\footnote{This discussion for model building with three family of SM fermions also applies to the generalized gauge symmetries $SU(4)_C\times SU(6)_L\times SU(2)_R$ and $SU(4)_C\times SU(2)_L\times SU(6)_R$.}.

Similar as for the standard D-brane construction, to satisfy the $I_{ac'} =0 $ condition, the stack $a$ D6-branes are constructed to be parallel to the orientifold ($\Omega R$) image $c'$ of the $c$ stack of D6-branes along at least one tow-torus. For which, we choose this to be the third two-torus in our convention and th open strings stretch between the $a$ and $c'$ stacks of D6-branes. 
When the minimal distance square $Z^2_{(ac')}$ (in $1/M_s$ units) between these two stacks 
on the third two-torus is small, the minimal length squared of the stretched string is small. The light scalars with squared-masses $Z^2_{(ab')}/(4\pi^2 \alpha')$ arise from the NS sector, while the light fermions with the same masses arise from R sector as discussed in~\cite{Uranga,imr,LLG3}. These scalars and fermions form four-dimensional $N=2$ hypermultiplets. 
One obtains $I_{ac'}^{(2)}$ number of vector-like pairs for the chiral superfields with quantum numbers $({\bf {\bar 4}, 1, 2})$ and $({\bf 4, 1, 2})$. 
These vector-like particles contribute as the Higgs fields  breaking the Pati-Salam gauge symmetry down to the SM gauge symmetry, with four-dimensional $N=1$ supersymmetry preserved. 
In particular, these fields are massless under the condition $Z^2_{(ac')}=0$. 
Note that the model with intersection numbers $I_{ac}=0$ and $I_{ac'}=-3$ are equivalent to the models with $I_{ac}=-3$ and $I_{ac'}=0$ under  the symmetry transformation $c\leftrightarrow c'$.

From the phenomenology aspect,  we now briefly review the procedures to break our generalized Pati-Salam gauge symmetry to the SM via realizing the Pati-Salam models. Firstly, we take models  with gauge symmetry $U(4)\times U(6)_L\times U(2)_R\times USp(2)$ and $U(4)\times U(2)_L\times U(6)_R\times USp(2)$ as example to discuss about the Higgs mechanism to break the $U(6)$ gauge to $U(2)$ and thus to discuss the resulting Pati-Salam models' phenomenology. Consider an $U(6)$ gauge theory with a scalar field in the adjoint representation. By taking proper orientations commute with the $U(6)$ generators, we can break $U(6)$ spontaneously to $U(2)\times U(2)\times U(2)$, and then to  $U(2)\times U(2)$ and in the end to  $U(2)$. In the following, we will show the orientation matrix and the masses of the massive bosons.
Firstly, we take the orientation for $U(6)$ scalar field act on the vacuum expectation value $\Phi_0$ as 
\begin{equation}
{\bf \Phi}_0={\bf \phi_{U(6)}}\, .
\begin{pmatrix}
1 \,& 0 \, & 0 \,&0\, &0 \,&0\\
0 \, & 1  \,& 0 &0 \,&0 \,&0 \\
0 \,& 0  \,& 2 \,&0 \,&0 \,&0 \\
0 \,& 0  \,& 0 \,&2 \,& 0 \,&0 \\
0 \,& 0  \,& 0 \,&0 \,& -3&0 \\
0 \,& 0  \,& 0 \,&0 \,& 0 &-3 
\end{pmatrix}\, ,
\end{equation} 
we found that the $U(6)$ gauge symmetry spontaneously to $U(2)\times U(2)\times U(2)$ by checking the commutator of the orientation and the $U(6)$ generators. As follows, to simplify the following discussion, we construct the $U(2)\times U(2)\times U(2)$ generators with Pauli matrix for each $U(2)$ and apply the orientation matrix in the manner of 
\begin{equation}
{\bf \Phi}_0={\bf \phi_{U(2)\times U(2)\times U(2)}}\, .
\begin{pmatrix}
0 \,& 0 \, & V_{12} \,&0\, &0 \,&0\\
0 \, & 0  \,& 0 &V_{12} \,&0 \,&0 \\
V_{12} \,& 0  \,& 0 \,&0 \,&0 \,&0 \\
0 \,& V_{12}  \,& 0 \,&0 \,& 0 \,&0 \\
0 \,& 0  \,& 0 \,&0 \,& 0&0 \\
0 \,& 0  \,& 0 \,&0 \,& 0 &0 
\end{pmatrix}\, ,
\end{equation} 
we found that the above expectation value breaks $U(2)\times U(2)\times U(2)$ to $U(2)\times U(2)$ and leaves the gauge bosons corresponding to the $U(2)\times U(2)$ generators massless. By taking a third orientation to the $U(2)\times U(2)$ gauge field
\begin{equation}
{\bf \Phi}_0={\bf \phi_{U(2)\times U(2)}}\, .
\begin{pmatrix}
0 \,& 0 \, & 0 \,&0\, &V_{13} \,&0\\
0 \, & 0  \,& 0 &0 \,&0 \,&V_{13} \\
0 \,& 0  \,& 0 \,&0 \,&0 \,&0 \\
0 \,& 0  \,& 0 \,&0 \,& 0 \,&0 \\
V_{13} \,& 0  \,& 0 \,&0 \,& 0&0 \\
0 \,& V_{13}  \,& 0 \,&0 \,& 0 &0 
\end{pmatrix}\, ,
\end{equation} 
we break the $U(2)\times U(2)$ gauge symmetry to the $U(2)$ and acquire the masses 
\begin{equation}
m^2=(4g|\phi|)^2 \left(V_{12}^2+V_{13}^2 \pm \sqrt{V_{12}^4-V_{12}^2\, V_{13}^2 +V_{13}^4}\right).
\end{equation} 
For the models with gauge symmetries $SU(12)_C\times SU(2)_L\times SU(2)_R$, the Higgs mechanism for breaking $U(12) \rightarrow U(4)\times U(4)\times U(4) \rightarrow U(4)\times U(4) \rightarrow U(4)$ will follow the same procedure also.
By splitting the $a$ stack of D6-branes into $a_1$ and $a_2$ stacks  with 6 and 2 D6-branes, the $U(4)_C$ gauge symmetry breaks in to $U(3)_C \times U(1)$ as shown in~\cite{Chen:2007zu}. 
The gauge fields and three chiral multiplets in the adjoint representation of $SU(4)_C$ will be broken down to the adjoint representations of $SU(3)_C$ as well as the gauge field and three singlets of $U(1)_{B-L}$. 
We note the number of symmetric  representations for $SU(4)_C$  as $n_{\Ysymm}^a$, while the anti-symmetric representations noted as $n_{\Yasymm}^a$. Similar convention applies to $SU(3)_C$, $SU(2)_L$, and $SU(2)_R$.
These chiral multiplets for $SU(4)_C$ are broken down to the $n_{\Ysymm}^a$ and $n_{\Yasymm}^a$ chiral multiplets in symmetric and anti-symmetric representations for $SU(3)_C$, and $n_{\Ysymm}^a$ chiral multiplets with $U(1)_{B-L}$ charge $\pm 2$.
Moreover, we have $I_{a_1 a'_2}$ new fields with quantum number $({\bf 3, -1})$ under $SU(3)_C\times U(1)_{B-L}$ from the open strings at the intersections of $a_1$ and $a_2'$ stacks of D6-branes, while the other particle spectrum stay the same.  
The anomaly free gauge symmetries $SU(3)_C\times U(1)_{B-L}$ arise from $a_1$ and $a_2$ stacks of D6-branes as the $SU(4)_C$ subgroup. 
To break the $U(2)_R$ gauge symmetry,  we split the $c$ stack of D6-branes into $c_1$ and $c_2$ stacks, and each with two D6-branes. 
The gauge fields and three chiral multiplets in the adjoint representation of $SU(2)_R$ break down to the gauge field and three singlets of $U(1)_{I_{3R}}$.
The $n_{\Ysymm}^c$ chiral multiplets in the symmetric representation of $SU(2)_R$ break down to the $n_{\Ysymm}^c$ chiral multiplets with $U(1)_{I_{3R}}$ charge, while the $n_{\Yasymm}^c$ chiral multiplets in anti-symmetric representation $SU(2)_R$ vanish. 
Arising from the open strings at the intersections of $c_1$ and $c_2'$ stacks of D6-brane, there are $I_{c_1 c'_2}$ new fields that are neutral under $U(1)_{I_{3R}}$. As follows, the anomaly free gauge symmetry from $c_1$ and $c_2$ stacks of D6-branes is $U(1)_{I_{3R}}$, the $SU(2)_R$ Cartan subgroup.
One obtains the  $SU(3)_C\times SU(2)_L\times U(1)_{B-L} \times U(1)_{I_{3R}}$ gauge symmetry the above D6-brane splittings. 
To break the gauge symmetry further to SM gauge symmetry, we consider the minimal distance square $Z^2_{(a_2 c_1')}$ to be small, and obtain $I_{a_2 c_1'}^{(2)}$  pairs of chiral multiplets with quantum numbers  $({\bf { 1}, 1, -1, 1/2})$ and $({\bf { 1}, 1, 1, -1/2})$ 
under $SU(3)_C\times SU(2)_L\times U(1)_{B-L} \times U(1)_{I_{3R}}$.  
Light open string  stretches between the $a_2$ and $c_1'$ stacks of D6-branes, and 
arises vector-like particles can then break the $SU(3)_C\times SU(2)_L\times U(1)_{B-L} \times U(1)_{I_{3R}}$ gauge symmetry  down to the SM. In the meantime,  D- and F-flatness is kept as their quantum numbers are the same as those of the right-handed neutrino and its complex conjugate. 
%Note that they are massless when $Z^2_{(a_2c_1')}=0$.  
In summary, the complete chains for symmetry breaking of our generalized Pati-Salam Models read
\begin{eqnarray}
\left.
\begin{array}{rcl}
SU(12)\times SU(2)_L \times SU(2)_R\\
SU(4)\times SU(6)_L \times SU(2)_R\\
SU(4)\times SU(2)_L \times SU(6)_R
\end{array} \right\} &&
\overrightarrow{\;\rm Higgs \;Mechanism\;}\;  SU(4)\times SU(2)_L \times SU(2)_R \nonumber\\&&
\overrightarrow{\;a\rightarrow a_1+a_2\;}\;  SU(3)_C\times SU(2)_L
\times SU(2)_R \times U(1)_{B-L} \nonumber\\&&
 \overrightarrow{\; c\rightarrow c_1+c_2 \;} \; SU(3)_C\times SU(2)_L\times
U(1)_{I_{3R}}\times U(1)_{B-L} \nonumber\\&&
 \overrightarrow{\;\rm Higgs \;
Mechanism\;} \; SU(3)_C\times SU(2)_L\times U(1)_Y~.~\,
\end{eqnarray}
For more details of the dynamical supersymmetry breaking of Type IIA orientifolds with intersecting D6-branes, we refer to~\cite{CLW}. In which the filler branes carrying $USp$ gauge symmetries 
are confining, and allows for  supersymmetry breaking via gaugino condensation. However, in our construction, we generalize this to supersymmetry breaking via other mechanisms and not restrict to confining filler branes.
The gauge kinetic function for a generic stack $x$ of D6-branes takes the form of~\cite{CLW}
\begin{eqnarray}\label{kfunction}
f_x =  {\bf \textstyle{1\over 4}} \left[ n^1_x n^2_x n^3_x S -
(\sum_{i=1}^3 2^{-\beta_j-\beta_k}n^i_x l^j_x l^k_x U^i)  \right]
,\,
\end{eqnarray}
in which the real parts of dilaton $S$ and moduli $U^i$ are
\begin{eqnarray}
{\rm Re}(S) = \frac{M_s^3 R_1^{1} R_1^{2} R_1^{3} }{2\pi g_{s}}~,~\, \\
{\rm Re}(U^{i}) = {\rm Re}(S)~ \chi_j \chi_k~,~\,
\end{eqnarray}
where $i\neq j\neq k$,  and $g_s$ to be the string coupling. 
We note the gauge coupling constant associated with a stack $x$ is 
\begin{eqnarray}
\label{g2}
g_{D6_x}^{-2} &=& |\mathrm{Re}\,(f_x)|.
\end{eqnarray}
In our generalized construction, the holomorphic gauge kinetic functions for  $SU(12)_C$,  $SU(2)_L$ and $SU(2)_R$ are associated with stacks $a$,  $b$, and $c$, respectively. 
Recall that the holomorphic gauge kinetic function before our generalization is shown in~\cite{bkl, Chen:2007zu, Li:2019nvi, Blumenhagen:2003jy}, we have our holomorphic gauge kinetic function for $U(1)_Y$ as  linear combination of these for $SU(12)$ and $SU(2)_R$ in the form
\begin{equation}\label{frelation}
f_Y =  \frac{3}{5} \,( \frac{2}{3}\, f_{a} + f_{c} ).
\end{equation}
Due to the pair of $(n_x^i, l_x^i)$ in the gauge kinetic function Eq.~\ref{kfunction}, the kinetic function $f_a$ will have an overall factor of 3, and rescale the value of $U(1)_Y$ gauge kinetic functions in Eq.~\ref{frelation} for gauge couplings of MSSM.
By taking care of the common factor of 3 between the kinetic function of $SU(12)$ and $SU(4)$,  the tree-level MSSM gauge couplings take the form of
\begin{equation}
3 \,g^2_{a} = \alpha\, g^2_{b} = \beta\, \frac{5}{3}g^2_Y = \gamma\, \left[\pi e^{\phi_4}\right]
\end{equation}
where $g_a^2,  g^2_{b}$, and $\frac{5}{3}g^2_Y$ are the strong, weak and hypercharge gauge couplings respectively, where $\alpha, \beta, \gamma$ as the ratios between them. Moreover, the K\"ahler potential reads
\begin{eqnarray}
K=-{\rm ln}(S+ \bar S) - \sum_{I=1}^3 {\rm ln}(U^I +{\bar
U}^I).~\,
\end{eqnarray}
Three stacks of D6-branes\,(carrying $U(4)_C\times U(2)_L \times U(2)_R$ gauge symmetry) determine the complex structure moduli $\chi_1$, $\chi_2$ and $\chi_3$ 
due to the four-dimensional $N=1$ supersymmetry conditions, with only one independent modulus field. To stabilize the moduli, one usually construct the models with at least two $USp$ groups with negative $\beta$ functions which can be confined and then allow for gaugino condensations as was discussed in~\cite{Taylor,RBPJS,BDCCM}. However, we will later present models not only with two and more $USp$ groups, but also with only one $USp$ group having negative $\beta$ function and still realize three family of particles. In general, the one-loop beta function for the $2N^{(i)}$ filler branes\,( on top of $i$-th O6-plane and carry $USp(N^{(i)})$ group) represents by~\cite{Cvetic:2004ui}
\begin{eqnarray}
\beta_i^g&=&-3({N^{(i)}\over2}+1)+2 |I_{ai}|+
|I_{bi}| +  |I_{ci}|
+3({N^{(i)}\over2}-1)\nonumber\\
        &=&-6+2 |I_{ai}|+  |I_{bi}|+  |I_{ci}|~.~\,
\label{betafun}
\end{eqnarray}
When supersymmetry is broken via gaugino condensations on the condition of at least two confining gauge groups in the hidden sector, we may need to consider gauge mediation since gravity mediation is much smaller. Thus, the supersymmetry CP problem may be solved as well. In the models with only one confining $USp$ gauge groups, the supersymmetry need to be broken with alternative mechanisms rather than gaugino condensations.

As we found in~\cite{Li:2019nvi} that for a three-family supersymmetric Pati-Salam model, the corresponding new three-family supersymmetric Pati-Salam models by exchanging $b$- and $c$-stacks of D6-branes is not an equivalent model but leads to new gauge coupling behaviours. We will employ this method to improve the gauge couplings in our new model buildings and present the new models in the next subsection. 

\subsection{Generalized Supersymmetric Pati-Salam Models}

Based on the generalized construction as we presented above, now we show the new generalized Pati-Salam models with their exact warpping numbers. 
Differently with the standard constructed Pati-Salam Models such as in Ref.~\cite{Cvetic:2004ui, Li:2019nvi}, we introduce three stacks of D6-branes, $a$, $b$, and $c$  with three times of the wrapping numbers of D6-brane for one of the stacks. Thus the corrsponding gauge symmetries will be $U(12)_C \times U(2)_L \times U(2)_R$,   $U(4)_C  \times U(6)_L  \times U(2)_R$, or $U(4)_C \times U(6)_L \times U(2)_R$. 
%Unlike the strategy in Ref.~\cite{Cvetic:2004ui}, we do not restrict ourselves with  that at least two $USp$ groups in the hidden sector have negative $\beta$ functions, and instead we do a broader scanning without any constraint on the hidden sector.
% while negative $\beta$ functions required. 
To obtain the particle spectra with odd generations of the SM fermions, and satisfying the RR tadpole cancellation conditions, we again concentrate on the new scanning with only one tilted torus as was discussed in Ref.~\cite{Cvetic:2004ui, Li:2019nvi}. In our convention, we choose the third two-torus to be tilted and study the generalized Pati-Salam models in the following.

Now we present the representative models obtained from our generalized gauge construction in table \ref{U12-1} \ref{U12-2}  \ref{U6-R} \ref{U6-L} \ref{fourOplaneT}.  In the first column for each table,
we denote the gauge constructions of D6-branes as  $a$, $b$, and $c$ stacks, and 1, 2, 3, and 4 stacks for the filler branes along $\Omega R$, $\Omega R\omega$, $\Omega R\theta\omega$,  $\Omega R\theta$ orientifold planes representing the $USp(N)$ gauge symmetries.  
In the second column, $N$ represents the  number of D6-branes for each stack. When 24, 12, 12 appear here, it means there are three times of wrapping than the standard Pati-Salam models.
 Moreover, in the third column we present the wrapping numbers of all the D6-branes  and specify the third set of wrapping numbers as for the tilted two-torus. In the remaining right columns, we show the intersection numbers between different stacks, with $b'$ and $c'$ denote the $\Omega R$ images of $b$ and $c$, respectively. In addition, we also present the relation among the  moduli parameters imposed by the four-dimensional $N=1$ supersymmetry conditions,  and the one-loop $\beta$ functions ($\beta^g_i$) for the hidden sector gauge symmetries in the table. 
 In particular, we also give the MSSM gauge couplings in the caption of each model for checking the gauge coupling unification. 
 Note that here the MSSM gauge coupling refers to the gauge coupling after generalized gauge construction breaking, {\emph i.e. }$U(12) \rightarrow U(4)$, $U(6)_L \rightarrow U(2)_L$,  $U(6)_R \rightarrow U(2)_R$. 
%Note that here we we do not require  at least two confining hidden gauge sectors which are needed to realize the moduli stabilization and supersymmetry breaking via gaugino condensation, and instead we show the models with only one confining hidden gauge sector, such as Model \ref{U6-L}, \ref{U6-R}, \ref{U12-1}, \ref{U12-2}.
\begin{table}
	[htb] \footnotesize
	\renewcommand{\arraystretch}{1.0}
	\caption{D6-brane configurations and intersection numbers in Model \ref{U12-1}, and its MSSM gauge coupling relation is $g^2_a=\frac{631}{150}g^2_b=\frac{671}{90}g^2_c=\frac{3355}{1612}(\frac{5}{3}g^2_Y)=\frac{8\times 2^{1/4} \times 703^{3/4}}{225}\pi e^{\phi^4}$.} 
	\label{U12-1}
	\begin{center}
		\begin{tabular}{|c||c|c||c|c|c|c|c|c|c|}
			\hline
			\rm{model} \ref{U12-1} & \multicolumn{9}{c|}{$U(12)\times U(2)_L\times U(2)_R\times USp(2)$}\\
			\hline \hline \rm{stack} & $N$ & $(n^1,l^1)\times (n^2,l^2)\times
			(n^3,l^3)$ & $n_{\Ysymm}$& $n_{\Yasymm}$ & $b$ & $b'$ & $c$ & $c'$ & 3  \\
			\hline
			$a$&  24& $(-1,-1)\times (1,0)\times (-1,1)$ & 0 & 0  & 0 & 1 & -1 & 0  & 0 \\
			$b$&  4& $(-2,1)\times (3,-1)\times (1,-1)$ & 2 & 22  & - & - & 8 & 0  & 2  \\
			$c$&  4& $(2,3)\times (-2,1)\times (-1,-1)$ & 13 & 35  & - & - & - & -  & -2\\
			\hline
			3&   2& $(0,-1)\times (1,0)\times (0,2)$& \multicolumn{7}{c|}{$X_A =\frac{19}{2} X_B = \frac{38}{37}X_C =\frac{19}{2}X_D$}\\
			&   &  & \multicolumn{7}{c|}{$\beta^g_2=-2$}\\
			&      &	                                                   & \multicolumn{7}{c|}{$\chi_1=\sqrt{\frac{38}{37}}, \chi_2=\frac{\sqrt{1406}}{4}, \chi_3=2\sqrt{\frac{38}{37}}$}\\               
			\hline
		\end{tabular}
	\end{center}
\end{table}

\begin{table}
	[htb] \footnotesize
	\renewcommand{\arraystretch}{1.0}
	\caption{D6-brane configurations and intersection numbers in Model \ref{U12-2}, and its MSSM gauge coupling relation is $g^2_a=\frac{671}{90}g^2_b=\frac{631}{150}g^2_c=\frac{3155}{1712}(\frac{5}{3}g^2_Y)=\frac{8\times 2^{1/4} \times 703^{3/4}}{225}\pi e^{\phi^4}$.} 
	\label{U12-2}
	\begin{center}
		\begin{tabular}{|c||c|c||c|c|c|c|c|c|c|}
			\hline
			\rm{model} \ref{U12-2} & \multicolumn{9}{c|}{$U(12)\times U(2)_L\times U(2)_R\times USp(2)$}\\
			\hline \hline \rm{stack} & $N$ & $(n^1,l^1)\times (n^2,l^2)\times
			(n^3,l^3)$ & $n_{\Ysymm}$& $n_{\Yasymm}$ & $b$ & $b'$ & $c$ & $c'$ & 1  \\
			\hline
			$a$&  24& $(1,1)\times (1,0)\times (1,-1)$ & 0 & 0  & 0 & 1 & 0 & -1  & 0 \\
			$b$&  4& $(3,-2)\times (2,-1)\times (-1,1)$ & 13 & 35  & - & - & - & -8  & -2\\
			$c$&  4& $(-1,2)\times (-3,-1)\times (1,-1)$ & -2 & -22  & - & - & 0 & -  & -2  \\
			\hline
			1&   2& $(1,0)\times (1,0)\times (2,0)$& \multicolumn{7}{c|}{$X_A =\frac{37}{4} X_B = \frac{37}{38}X_C =\frac{37}{4}X_D$}\\
			&   &  & \multicolumn{7}{c|}{$\beta^g_2=-2$}\\
			&      &	                                                   & \multicolumn{7}{c|}{$\chi_1=\sqrt{\frac{37}{38}}, \chi_2=\frac{\sqrt{1406}}{4}, \chi_3=\sqrt{\frac{74}{19}}$}\\               
			\hline
		\end{tabular}
	\end{center}
\end{table}	

\begin{table}
	[htb] \footnotesize
	\renewcommand{\arraystretch}{1.0}
	\caption{D6-brane configurations and intersection numbers in Model \ref{U6-R}, and its MSSM gauge coupling relation is $g^2_a=\frac{7}{6}g^2_b=\frac{3}{2}g^2_c=\frac{5}{4}(\frac{5}{3}g^2_Y)=\frac{8\times2^{1/4}}{3}\pi e^{\phi^4}$.} \label{U6-R}
	\begin{center}
		\begin{tabular}{|c||c|c||c|c|c|c|c|c|c|}
			\hline
			\rm{model} \ref{U6-R} & \multicolumn{9}{c|}{$U(4)\times U(2)_L\times U(6)_R\times USp(2)$}\\
			\hline \hline \rm{stack} & $N$ & $(n^1,l^1)\times (n^2,l^2)\times
			(n^3,l^3)$ & $n_{\Ysymm}$& $n_{\Yasymm}$ & $b$ & $b'$ & $c$ & $c'$& 1  \\
			\hline
			$a$&  8& $(-1,0)\times (1,1)\times (-1,1)$ & 0 & 0  & 0 & 3 & -1 & 0 & 0  \\
			$b$&  4& $(-1,-1)\times (1,2)\times (-1,1)$ & 0 & 8  & - & - & 0 & 0 & 2  \\
			$c$&  12& $(0,-1)\times (1,2)\times (1,1)$ & -1 & 1  & - & - & - & - & 2 \\
			\hline
			1&   2& $(1,0)\times (1,0)\times (2,0)$& \multicolumn{7}{c|}{$X_A =\frac{1}{2} X_B = \frac{1}{4}X_C =\frac{1}{4}X_D$}\\
			&   &  & \multicolumn{7}{c|}{$\beta^g_1=-2$}\\
			&      &	                                                   & \multicolumn{7}{c|}{$\chi_1=\frac{1}{2\sqrt{2}}, \chi_2=\frac{1}{\sqrt{2}}, \chi_3=\sqrt{2}$}\\               
			\hline
		\end{tabular}
	\end{center}
\end{table}

\begin{table}
	[htb] \footnotesize
	\renewcommand{\arraystretch}{1.0}
	\caption{D6-brane configurations and intersection numbers in Model \ref{U6-L}, and its MSSM gauge coupling relation is $g^2_a=\frac{3}{2}g^2_b=\frac{7}{6}g^2_c=\frac{35}{32}(\frac{5}{3}g^2_Y)=\frac{8\times 2^{1/4}}{3}\pi e^{\phi^4}$.} \label{U6-L}
	\begin{center}
		\begin{tabular}{|c||c|c||c|c|c|c|c|c|c|}
			\hline
			\rm{model} \ref{U6-L} & \multicolumn{9}{c|}{$U(4)\times U(6)_L\times U(2)_R\times USp(2)$}\\
			\hline \hline \rm{stack} & $N$ & $(n^1,l^1)\times (n^2,l^2)\times
			(n^3,l^3)$ & $n_{\Ysymm}$& $n_{\Yasymm}$ & $b$ & $b'$ & $c$ & $c'$& 1  \\
			\hline
			$a$&  8& $(-1,1)\times (1,0)\times (-1,-1)$ & 0 & 0  & 0 & 1 & 0 & -3 & 0 \\
			$b$&  12& $(-1,-2)\times (0,1)\times (1,1)$ & -1 & 1  & - & - & 0 & 0 & 2  \\
			$c$&  4& $(-1,2)\times (1,-1)\times (-1,-1)$ & 0 & -8  & - & - & - & - & -2 \\
			\hline
			1&   2& $(1,0)\times (1,0)\times (2,0)$& \multicolumn{7}{c|}{$X_A =\frac{1}{4} X_B = \frac{1}{2}X_C =\frac{1}{4}X_D$}\\
			&   &  & \multicolumn{7}{c|}{$\beta^g_1=-2$}\\
			&      &	                                                   & \multicolumn{7}{c|}{$\chi_1=\frac{1}{\sqrt{2}}, \chi_2=\frac{1}{2\sqrt{2}}, \chi_3=\sqrt{2}$}\\               
			\hline
		\end{tabular}
	\end{center}
\end{table}

The Higgs particles in Models \ref{U12-1}, \ref{U6-R} arise from $N=2$ subsectors at the intersections of b- and  $c'$-stacks of D6-branes, while the Higgs particles in Models \ref{U12-2}, \ref{U6-L} arise from $N=2$ subsectors at the intersections of b- and c-stacks of D6-branes. For example, there exist 20 exotic Higgs-like particles in Models \ref{U12-2} from $N=2$ subsectors at the intersections of b- and  c-stacks of D6-branes.
We show that while there are $24$ number of D6-brane constructed in the a-stack, the gauge group yields to $U(12)\times U(2)_L\times U(2)_R$. With gauge breaking,  $U(12) \rightarrow U(4)$, we obtain the standard Pati-Salam model gauge  $U(4)\times U(2)_L\times U(2)_R$ and its MSSM gauge couplings after generalized gauge construction breaking. Model \ref{U12-1} and \ref{U12-2} are with their b- and c-stacks of brane swapped, and as expected, the gauge coupling get rescaled and refined. However, for models like Model \ref{fourOplaneT}, the gauge coupling from the b- and c-stacks of brane $g^2_b=g^2_c$, such that the b- and c-stacks swapping would not change the MSSM gauge couplings\footnote{Note that from Eq.~\eqref{frelation}, it is easy to check that when one model has $g^2_Y=g^2_a$, by applying b- and  c-stacks of brane swapping, the new model is with $g^2_a=g^2_b$, and vice versa.  This we have shown with detailed examples in \cite{Li:2019nvi}. In principle, by replacing the holomorphic gauge function $f_c$ with $f_b$ in Eq.~\eqref{frelation}, one can compute and predict all the gauge couplings behaviours after b- and  c-stacks of brane swapping without reconstructing the dual model.}.

Now we move on to the examples of generalized gauge construction for $U(4)\times U(2)_L\times U(6)_R\times USp(2)$ and $U(4)\times U(6)_L\times U(2)_R\times USp(2)$ with gauge breaking  $U(6)_R \rightarrow U(2)_R$ and $U(6)_L \rightarrow U(2)_L$ as shown in Model \ref{U6-R} and  \ref{U6-L}. We observe that the generalized gauge construction got shifted from $U(6)_R$ to $U(6)_L$ and the $U(1)_Y$ gauge coupling $\frac{5}{3}g^2_Y$ got rescaled from $\frac{4}{5}g^2_a$ to  $\frac{32}{35}g^2_a$ while $g^2_a$ remain the same.
Note that this construction is not simply swapping the b-stack and c-stack of D6-branes, but non-trivial T-dualities are performed to obtain models with three families of particles and tadpole cancellation conditions fulfilled. 
\begin{table}
	[htb] \footnotesize
	\renewcommand{\arraystretch}{1.0}
	\caption{D6-brane configurations and intersection numbers in Model \ref{fourOplaneT},
		and its MSSM gauge coupling relation is
		$g^2_a= 4\, g^2_b= 4\, g^2_c=\frac{20}{11}\,g^2_Y= {4}\sqrt{2} \, \pi \,e^{\phi^4}$.}
	\label{fourOplaneT}
	\begin{center}
		\begin{tabular}{|c||c|c||c|c|c|c|c|c|c|c|c|c|}
			\hline	 \rm{Model} \ref{fourOplaneT}&
			\multicolumn{12}{|c|}{$U(4)\times U(2)_L\times U(2)_R\times USp(2)^4$}\\
			\hline \hline \rm{stack} & $N$ & $(n^1,l^1)\times (n^2,l^2)\times
			(n^3,l^3)$ & $n_{\Ysymm}$& $n_{\Yasymm}$ & $b$ & $b'$ & $c$ & $c'$& 1 & 2 & 3 & 4 \\
			\hline
			$a$&  24& $(1,-1)\times (0,-1)\times (-1,1)$ & 0 & 0  & 1 & 0 & 0 & -1 & -1 & 0 & 1 & 0  \\
			$b$&  4& $(1,-2)\times (1,-3)\times (1,1)$ & -2 & -22  & - & - & 0 & 0 & -6 & -2 &-3& 1   \\
			$c$&  4& $(2,1)\times (-1,3)\times (-1,1)$ & -2 & -22  & - & - & - & - & -3 & 1 & -6 & -2  \\
			\hline
			1&   2& $(1,0)\times (1,0)\times (2,0)$& \multicolumn{10}{c|}{$X_A = \frac{1}{9}\,X_B =X_C =\frac{1}{9}X_D$}\\
			2&   2& $(1,0)\times (0,-1)\times (0,2)$ & \multicolumn{10}{c|}{$\beta^g_1=5, \beta^g_2=-3, \beta^g_3=5, \beta^g_4=-3$}\\
			3&  2 & $(0,-1)\times (1,0)\times (0,2)$ & \multicolumn{10}{c|}{$\chi_1=1,\chi_2=\frac{1}{9}, \chi_3=2$}\\
			4&  2 & $(0,-1)\times (0,1)\times (2,0)$ & \multicolumn{10}{c|}{}\\
			\hline
		\end{tabular}
	\end{center}
\end{table}
For models with more than one orientifold plane in the generalized Pati-Salam models, with two confining gauge groups, there may exist the stable extrema with  moduli stabilization and supersymmetry breaking via gaugino condensations, which are very interesting from the phenomenological points of view. We show such example in Model  \ref{fourOplaneT} from generalized Pati-Salam construction. We note that for generalized construction with three times of wrapping at the a-stack of D-brane, there are only $USp(2)$ filler branes.
%However, as pointed out in Ref.~\cite{CLW}, the cosmological constants at these extrema are likely to be negative and close to the string scale, and thus the gaugino condensations in these models might not address the cosmological constant problem.

By employing a more powerful supervised scanning methods as we performed in \cite{Li:2019nvi}, we obtain models with wrapping numbers not only equal but also larger than $5$ which we shown in the Appendix. We expect that with supervised scanning methods, Models beyond the wrapping number $10$ can also be obtained in reasonable time period. 
	
\section{Preliminary Phenomenological Studies}
\label{sec:pheno}

In this section, we shall discuss the phenomenological features of our generalized Pati-Salam  models. 
The $\beta$ functions of $USp(2)$ groups are with at least one of them being negative.
For the model with two confining $USp(2)$ groups, we can break supersymmety via gaugino condensation, and decouple the exotic particles.
For the rest models with only one confining $USp(2)$ group, we need to address the modulus stabilization issue as well, which is generic for the models with one $USp$ group.

Resulting to the standard Pati-Salam models, now we can discuss the other spectrum as for the standard Pati-Salam models with gauge symmetry $SU(4)\times SU(2)_L \times SU(R)$ as we studied in \cite{Li:2019nvi}. 
We start with Models \ref{U12-1} and \ref{U12-2}, which are constructed with one $USp$ group. The gauge symmetry is $U(12)\times U(2)_L\times U(2)_R\times USp(2)$. We show the explicit spectrum  for Model \ref{U12-2} in Table \ref{u12-2}.
For this model, it is obvious that the Higgs multiplets therein is from the intersection of b- and c-stack of branes. 
The Models  \ref{U6-R} and \ref{U6-L} are constructed with gauge symmetries $U(4)\times U(6)_L\times U(2)_R\times USp(2)$ and $U(4)\times U(2)_L\times U(6)_R\times USp(2)$ respectively. We present the explicit spectrum  for Model \ref{U6-R} in Table \ref{u6-R}.
%Now we take Model \ref{U6-R} as example to show the spectrum for the models with $SU(6)$ gauge symmetry. 
For Model \ref{U6-R}, it is obvious that the Higgs multiplets therein is from the intersection of b- and $c'$-stack of branes. 
\begin{table}
	[htb] \footnotesize
	\renewcommand{\arraystretch}{1.0}
	\caption{The chiral spectrum in the open string sector for Model \ref{U12-2}. } 
	\label{u12-2}
	\begin{center}
		\begin{tabular}{|c||c||c|c|c||c|c|c|}\hline
			Model~\ref{U12-2} & $SU(12)\times SU(2)_L\times SU(2)_R \times USp(2)$
			& $Q_{12}$ & $Q_{2L}$ & $Q_{2R}$ & $Q_{em}$ & $B-L$ & Field \\
	%		& $\times USp(4)$ & & & & & & \\
			\hline\hline
			$ab'$ & $1 \times (12, 2,1,1)$ & 1 & 1 & 0  & $-\frac 13,\; \frac 23,\;-1,\; 0$ & $\frac 13,\;-1$ & $Q_L, L_L$\\
			$ac'$ & $1 \times (\overline{12},1,\overline{2},1)$ & -1 & 0 & -1   & $\frac 13,\; -\frac 23,\;1,\; 0$ & $-\frac 13,\;1$ & $Q_R, L_R$\\
			$bc'$ & $8 \times(1,\overline{2}, \overline{2},1)$ & 0 & -1 & -1   & $-1,\;0,\;0,\;1$ & 0 & $H'$\\
			$b2$ & $2\times(1,\overline{2},1,2)$ & 0 & -1 & 0   & $\mp \frac 12$ & 0 & \\
  $c2$ & $2\times(1,1,\overline{2},2)$ & 0 & 0 & -1  & $\mp \frac 12$ & 0 & \\
			$b_{\Ysymm}$ & $13\times(1,3,1,1)$ & 0 & $2$ & 0   & $0,\pm 1$ & 0 & \\
			$b_{\overline{\Yasymm}}$ & $35\times(1, 1,1,1)$ & 0 & 2 & 0   & 0 & 0 & \\
			$c_{\overline{\Ysymm}}$ & $2\times(1,1,\overline{3},1)$ & 0 & 0 & -2   & $0,\pm 1$ & 0 & \\
			$c_{\Yasymm}$ & $22\times(1,1, {1},1)$ & 0 & 0 & -2   & 0 & 0 & \\
				\hline\hline
$bc$ & $20 \times (1,2,\overline{2},1)$ & 0 & 1 & -1   &  & &  \\
& $20 \times (1,\overline{2},2,1)$ &  0 & -1 &  1& $-1,\;0,\; 0,\; 1 $ &0 &$H_u^i, H_d^i$\\
	\hline
		\end{tabular}
	\end{center}
\end{table}

\begin{table}
	[htb] \footnotesize
	\renewcommand{\arraystretch}{1.0}
	\caption{The chiral spectrum in the open string sector for Model \ref{U6-R}. } 
	\label{u6-R}
	\begin{center}
		\begin{tabular}{|c||c||c|c|c||c|c|c|}\hline
			Model~\ref{U6-R} & $SU(4)\times SU(2)_L\times SU(6)_R \times USp(2)$
			& $Q_{4}$ & $Q_{2L}$ & $Q_{6R}$ & $Q_{em}$ & $B-L$ & Field \\
			%		& $\times USp(4)$ & & & & & & \\
			\hline\hline
			$ab'$ & $3 \times (4, 2,1,1)$ & 1 & 1 & 0  & $-\frac 13,\; \frac 23,\;-1,\; 0$ & $\frac 13,\;-1$ & $Q_L, L_L$\\
			$ac$ & $1 \times (\overline{4},1,6,1)$ & -1 & 0 & 1   & $\frac 13,\; -\frac 23,\;1,\; 0$ & $-\frac 13,\;1$ & $Q_R, L_R$\\
			$b2$ & $2\times(1,2,1,\overline{2})$ & 0 & 1 & 0   & $\pm \frac 12$ & 0 & \\
			$c2$ & $2\times(1,1,6,\overline{2})$ & 0 & 0 & 1  & $\pm \frac 12$ & 0 & \\
			$b_{\overline{\Yasymm}}$ & $8\times(1, 1,1,1)$ & 0 & 2 & 0   & 0 & 0 & \\
			$c_{\overline{\Ysymm}}$ & $1\times(1,1,\overline{21},1)$ & 0 & 0 & -2   & $0,\pm 1$ & 0 & \\
			$c_{\Yasymm}$ & $1\times(1,1,15,1)$ & 0 & 0 & 2   & 0 & 0 & \\
			\hline\hline
			$bc'$ & $4 \times (1,2,6,1)$ & 0 & 1 & 1   &  & & \\
			& $4 \times (1,\overline{2},\overline{6},1)$  & 0 & -1 & -1   & $-1,\;0,\; 0,\;1$ &0 & $H_u^i, H_d^i$\\
			\hline
		\end{tabular}
	\end{center}
\end{table}

Now we take Model \ref{U6-R} as examples to show explicitly the new composite states  formed due to the strong forces from hidden sector. We present the confined particle spectrum in Table \ref{Composite Particles U6-R}. Because it has 
one confining gauge group $USp(2)$ with two charged intersections. Therefore, besides self-confinement, the mixed-confinement between different intersections is also possible, which yields the chiral supermultiplets $(1,2, 6,1)$.
All the models we presented contain exotic particles that are charged under the hidden gauge groups. The strong coupling dynamics in the hidden sector at certain intermediate scale might provide a mechanism for all these particles to form bound states or composite particles. These are compatible with anomaly cancellation conditions, such that we do not have extra anomaly introduced. Moreover, similar to the quark condensation in QCD, these particles will only be charged under the SM gauge symmetry~\cite{CLS1}. 
In general, these $USp$ groups have two kinds of neutral bound states. The first one is the pseudo inner product of two fundamental representations generated by decomposing the rank two anti-symmetric representation. This can be considered as the reminiscent of a meson formed by the inner product of one pair of fundamental and anti-fundamental representations of $SU(3)_C$ in QCD.  This applies to our models.
The second kind is with rank $2N$ anti-symmetric representation of $USp(2N)$ group for $N\geq 2$, which is an $USp(2N)$ singlet and somewhat similar to a  baryon, as a rank three anti-symmetric representation of $SU(3)_C$  in QCD. 
Recall that our generalized models are with confining groups $USp(2)$ only in the hidden sector, such that the second is not covered in the obtained models so far.  

\begin{table}
	[htb] \footnotesize
	\renewcommand{\arraystretch}{1.0}
	\caption{Composite particle spectrum for Model \ref{U6-R}.}
	\label{Composite Particles U6-R}
	\begin{center}
		\begin{tabular}{|c|c||c|c|}\hline
			\multicolumn{2}{|c||}{Model \ref{U6-R}} &
			\multicolumn{2}{c|}{$SU(4)\times SU(2)_L\times SU(6)_R \times USp(2)$} \\
			\hline Confining Force & Intersection & Exotic Particle
			Spectrum & Confined Particle Spectrum \\
			\hline\hline
			$USp(2)_1$ &$b2$ & $2\times(1, 2,1, \overline{2})$ & $4\times(1,1,1,1)$, $4\times(1,3,1,1)$, $4\times(1,2, 6,1)$\\
			&$c2$ & $2\times(1,1,6,\overline{2})$ & $4\times(1,1,15,1)$, $4\times(1,1,21,1)$\\
			\hline
		\end{tabular}
	\end{center}
\end{table}

\begin{table}
	[htb] \footnotesize
	\renewcommand{\arraystretch}{1.0}
	\caption{The chiral spectrum in the open string sector for  Model \ref{fourOplaneT}.} \label{spectrum fourOplaneT}
	\begin{center}
		\begin{tabular}{|c||c||c|c|c||c|c|c|}\hline
			Model \ref{fourOplaneT} & $SU(12)\times SU(2)_L\times SU(2)_R \times USp(2)^4$
			& $Q_{12}$ & $Q_{2L}$ & $Q_{2R}$ & $Q_{em}$ & $B-L$ & Field \\
			\hline\hline
			$ab$ & $1 \times (12,\overline{2},1,1,1,1,1)$ & 1 & -1 & 0  & $-\frac 13,\; \frac 23,\;-1,\; 0$ & $\frac 13,\;-1$ & $Q_L, L_L$\\
			$ac'$ & $1\times (\overline{12},1,\overline{2},1,1,1,1)$ & -1 & 0 & -1  & $\frac 13,\; -\frac 23,\;1,\; 0$ & $-\frac 13,\;1$ & $Q_R, L_R$\\
			$a1$ & $1\times (\overline{12},1,1,2,1,1,1)$ & -1& 0 & 0 & $-\frac 16,\;\frac 12$ & $-\frac 13,\;1$ & \\
			$a3$ & $1\times (12,1,1,1,1,\overline{2},1)$ & 1 & 0 & 0   & $\frac 16,\;-\frac 12$ & $\frac 13,\;-1$ & \\
    $b1$ & $6\times(1,\overline{2},1,2,1,1,1)$ & 0 & -1 & 0   & $\mp \frac 12$ & 0 & \\
			$b2$ & $2\times(1,\overline{2},1,1,2,1,1)$ & 0 & -1 & 0   & $\mp \frac 12$ & 0 & \\
	$b3$ & $3\times(1,\overline{2},1,1,1,2,1)$ & 0 & -1 & 0   & $\mp \frac 12$ & 0 & \\
			$b4$ & $1\times(1,2,1,1,1,1,\overline{2})$ & 0 & 1 & 0   & $\pm \frac 12$ & 0 & \\
			$c1$ & $3\times(1,1,\overline{2},2,1,1,1)$ & 0 & 0 & -1   & $\mp \frac 12$ & 0 & \\
	$c2$ & $1\times(1,1,2,1,\overline{2},1,1)$ & 0 & 0 & 1   & $\pm \frac 12$ & 0 & \\
			$c3$ & $6\times(1,1,\overline{2},1,1,2,1)$ & 0 & 0 & -1   & $\mp \frac 12$ & 0 & \\
	$c4$ & $2\times(1,1,\overline{2},1,1,1,2)$ & 0 & 0 & -1   & $\mp \frac 12$ & 0 & \\
			$b_{\Ysymm}$ & $2\times(1,\overline{3},1,1,1,1,1)$ & 0 & -2 & 0   & $0,\pm 1$ & 0 & \\
			$b_{\overline{\Yasymm}}$ & $22\times(1,1,1,1,1,1,1)$ & 0 & 2 & 0   & 0 & 0 & \\
			$c_{\overline{\Ysymm}}$ & $2\times(1,1,\overline{3},1,1,1,1)$ & 0 & 0 & -2   & $0,\pm 1$ & 0 & \\
			$c_{\Yasymm}$ & $22\times(1,1,1,1,1,1,1)$ & 0 & 0 & 2   & 0 & 0 & \\
						\hline\hline
			$bc'$ & $18 \times (1,2,2,1,1,1,1)$ & 0 & 1 & 1   &  & & \\
			& $18 \times (1,\overline{2},\overline{2},1,1,1,1)$  & 0 & -1 & -1   & $-1,\;0,\; 0,\;1$ &0 & $H_u^i, H_d^i$\\
			\hline
		\end{tabular}
	\end{center}
\end{table}
%For Model \ref{fourOplaneT}, there are two confining $USp(N)$ gauge groups, and thus  the non-perturbative effective superpotential can be generated via gaugino condensations.  The ground state, which is determined by the minimization of this supergravity potential, can stabilize the dilaton and complex structure toroidal moduli, and breaks supersymmetry~\cite{CLW}.  
For Model \ref{fourOplaneT}, there are two confining $USp(N)$ gauge groups, a general analysis of the non-perturbative superpotential with tree-level gauge couplings can be performed, and it was shown that there can exist extrema with the stabilizations of dilaton and complex structure moduli~\cite{CLW}. 
However, these extrema of such model might be saddle points and thus do not break supersymmetry. For further investigation, if the models have three or four confining $USp(N)$ gauge groups,  the non-perturbative superpotential allows for the  moduli stabilization and supersymmetry breaking at the stable extremum in general~\cite{CLW}. 
As it is shown in Table \ref{spectrum fourOplaneT}, there are many more intersections with the filler branes for the generalized Model \ref{fourOplaneT} with four $USp(2)$ gauge groups constructed, which result in many more composited states appear. The confined particles spectra  formed due to the strong forces from hidden sector are tabulated in Table \ref{Composite Particles fourOplaneT}.
In more detail, Model \ref{fourOplaneT} has four confining gauge groups with $USp(2)$ groups. Thereinto, both $USp(2)_1$ and $USp(2)_3$ have three charged intersections, while $USp(2)_2$
and $USp(2)_4$ have two charged intersections. So for them, besides self-confinement, the mixed-confinement between different intersections become many more, which yields the chiral
supermultiplets $(\overline{12},\overline{2},1,1,1,1,1)$, $(\overline{12},1,\overline{2},1,1,1,1)$, $(1,\overline{2},\overline{2},1,1,1,1)$, $(1,\overline{2},2,1,1,1,1)$, $(12,\overline{2},1,1,1,1,1)$, $(12,1,\overline{2},1,1,1,1)$, $(1,\overline{2},\overline{2},1,1,1,1)$, and $(1,2,\overline{2},1,1,1,1)$.

\begin{table}
	[htb] \footnotesize
	\renewcommand{\arraystretch}{1.0}
	\caption{Composite particle spectrum of Model \ref{fourOplaneT}.}
	\label{Composite Particles fourOplaneT}
	\begin{center}
		\begin{tabular}{|c|c||c|c|}\hline
			\multicolumn{2}{|c||}{Model \ref{fourOplaneT}} &
			\multicolumn{2}{c|}{$SU(12)\times SU(2)_L\times SU(2)_R \times
				USp(2)^4$} \\
			\hline Confining Force & Intersection & Exotic Particle
			Spectrum & Confined Particle Spectrum \\
			\hline\hline
			$USp(2)_1$ &$a1$ & $1\times (\overline{12},1,1,2,1,1,1)$ & $1\times (\overline{66},1,1,1,1,1,1)$, $1\times (\overline{78},1,1,1,1,1,1)$,
			 $1\times(\overline{12},\overline{2},1,1,1,1,1)$\\
&$b1$ & $1\times(1,\overline{2},1,2,1,1,1)$ & $1\times(1,1,1,1,1,1,1)$ , $1\times(1,\overline{3},1,1,1,1,1)$ ,
 $1\times(\overline{12},1,\overline{2},1,1,1,1)$\\
			&$c1$ & $1\times(1,1,\overline{2},2,1,1,1)$ & $1\times(1,1,1,1,1,1,1)$, $1\times(1,1,\overline{3},1,1,1,1)$, 
			$1\times(1,\overline{2},\overline{2},1,1,1,1)$\\
			\hline
			$USp(2)_2$ &$b2$ & $1\times (1,\overline{2},1,1,2,1,1)$ & $1\times (1,1,1,1,1,1,1)$, $1\times (1,\overline{3},1,1,1,1,1)$,
			 $1\times(1,\overline{2},2,1,1,1,1)$\\
			&$c2$ & $1\times(1,1,2,1,\overline{2},1,1)$ &  $1\times(1,1,1,1,1,1,1)$, $1\times(1,1,3,1,1,1,1)$\\
			\hline
			$USp(2)_3$ &$a3$ & $3\times(12,1,1,1,1,\overline{2},1)$ &  $6\times(66,1,1,1,1,1,1)$, $6\times(78,1,1,1,1,1,1)$, $3\times(12,\overline{2},1,1,1,1,1)$\\
&$b3$ & $1\times(1,\overline{2},1,1,1,2,1)$ &  $1\times(1,1,1,1,1,1,1)$,
$1\times(1,\overline{3},1,1,1,1,1)$,
 $3\times(12,1,\overline{2},1,1,1,1)$\\
&$c3$ & $1\times(1,1,\overline{2},1,1,2,1)$ &  $1\times(1,1,1,1,1,1,1)$,
$1\times(1,1,\overline{3},1,1,1,1)$,
 $1\times(1,\overline{2},\overline{2},1,1,1,1)$\\
			\hline
			$USp(2)_4$ &$b4$ & $3\times(1,2,1,1,1,1,\overline{2})$ & $6\times(1,1,1,1,1,1,1)$, $6\times(1,3,1,1,1,1,1)$, $3\times(1,2,\overline{2},1,1,1,1)$\\
			&$c4$ & $1\times(1,1,\overline{2},1,1,1,2)$ &  $1\times(1,1,1,1,1,1,1)$, $1\times(1,1, \overline{3},1,1,1,1)$\\
			\hline
		\end{tabular}
	\end{center}
\end{table}

\section{Discussions and Conclusion}
\label{sec:conclusion}

We generalized the construction of three-family $N=1$ supersymmetric Pati-Salam models from Type IIA orientifolds on $\IT^6/(\IZ_2\times \IZ_2)$ with intersecting D6-branes, where the $SU(12)_C\times SU(2)_L \times SU(2)_R$, $SU(4)_C\times SU(6)_L \times SU(2)_R$ or $SU(4)_C\times SU(2)_L \times SU(6)_R$ gauge symmetries  arise from the stacks of D6-branes with  $U(n)$ gauge symmetries. Firstly, via Higgs mechanism we can break the  generalized Pati-Salam gauge symmetry $U(12) \rightarrow U(4)\times U(4)\times U(4) \rightarrow U(4)\times U(4) \rightarrow U(4)$  and  $U(6) \rightarrow U(2)\times U(2)\times U(2) \rightarrow U(2)\times U(2) \rightarrow U(2)$, with new massive bosons obtained in this procedure, and resulting in standard Pati-Salam gauge symmetry. 
Taking the gauge group $U(6)$ breaking to $U(2)$ as example, we studied the gauge symmetry breaking in details from the generalized models to the standard Pati-Salam models, and computed the masses of the gauge bosons in this Higgs mechanism.  
The Pati-Salam gauge symmetry can then be broken down to the  $SU(3)_C\times SU(2)_L\times U(1)_{B-L} \times U(1)_{I_{3R}}$ via D6-brane splittings, and further down to the SM gauge symmetry via the D- and F-flatness preserving Higgs mechanism in which Higgs fields are the massless open string states from a specific $N=2$ subsector. 
Moreover,  Model \ref{U12-1} with b-stack and c-stack of D6-branes swapped leads to Model \ref{U12-2} with $U(1)_Y$ and $SU(4)_C$ gauge couplings being closer to unification at the string scale. Model \ref{U6-L} and \ref{U6-R} with the b-stack and c-stack of D6-branes swapping also improved on the coupling unification aspects. Moreover, we observed that in the generalized Pati-Salam construction, there appear many more intersections with the filler branes in Model \ref{fourOplaneT}, and many more confine particles present.
%%%%%%%%
In addition, the models with large wrapping numbers equal and larger than $5$ can also be generalized in our generalized construction. 

As an outlook, we would like to address that another interesting scenario worthwhile to try is to construct the $SU(2)_L$ and/or $SU(2)_R$ gauge symmetries from filler branes, namely, $SU(2)_{L,R}=USp(2)_{L,R}$. As follows, the number of the SM Higgs doublet pairs might be decreased. In this case, one usually shall not construct the $SU(2)_{L,R}$ gauge symmetries from the splittings of higher rank $USp(N)$ ($N\ge 4$) branes, as it in general leads to even number of families, and the absolute value for one wrapping number of $U(4)$ branes larger than 2 cannot be avoided. This makes it difficult due to the tadpole cancellation conditions for model building and calls for more powerful  scanning methods for model buildings.

\section*{Acknowledgments} 

TL and AM are supported  by the Projects 11847612 and 11875062 from the National Natural Science Foundation of China, the Key Research Program of Frontier Science, CAS. RS is supported by the National Thousand Young Talents Program of China, the China Postdoctoral Science Foundation Grant 2018M631436, and the LMU Munich's Institutional Strategy LMUexcellent within the framework of the German Excellence Initiative. We would like to thank Chi-Ming Chang and Xiaoyong Chu for useful discussions. RS would also like to acknowledge Ludwig Maximilian University of Munich, Max Planck Institute for Physics, and the Abdus Salam International Centre for Theoretical Physics~(ICTP) for their hospitalities where part of this work was carried out.

%\newpage
%%%%%%%%%%%%%%%%%%%%%%%%%%%%%%%%%%%%%%%%%%%%%%%%%%%%%%%%
%                 MODEL LISTINGS                       %
%%%%%%%%%%%%%%%%%%%%%%%%%%%%%%%%%%%%%%%%%%%%%%%%%%%%%%%%

\begin{appendix}

\begin{center}
	\Large{\bf Appendix: Tables for Supersymmetric Pati-Salam Models}
\end{center}

In this Appendix,  we tabulate $8$ representative models obtained from our generalized gauge construction.  In the first column for each table,
we denote the gauge constructions of D6-branes as  $a$, $b$, and $c$ stacks, respectively. We denote 1, 2, 3, and 4 stacks for the filler branes along $\Omega R$, $\Omega R\omega$, $\Omega R\theta\omega$, and $\Omega R\theta$ orientifold planes respectively, which represent the $USp(N)$ gauge symmetries.  
In the second column, $N$ is the number of D6-branes for each stack. Moreover, in the third column we present the wrapping numbers of the various D6-branes  and specify the third set of wrapping numbers for the tilted two-torus. In the remaining right columns, we show the intersection numbers between different stacks, where $b'$ and $c'$ denote the $\Omega R$ images of $b$ and $c$, respectively. In addition, we present the relation among the  moduli parameters imposed by the four-dimensional $N=1$ supersymmetry conditions,  and the one-loop $\beta$ functions ($\beta^g_i$) for the hidden sector gauge symmetries. 

In particular, we give the MSSM gauge couplings in the caption of	each model to check the gauge coupling unification. Note that here the MSSM gauge coupling refers to the gauge coupling after generalized gauge construction breaking, i.e. $U(12) \rightarrow U(4)$, $U(6)_L \rightarrow U(2)_L$,  $U(6)_R \rightarrow U(2)_R$. Model \ref{model2} and  \ref{model11},   \ref{model3}  and \ref{model15}, \ref{model5} and \ref{model9}, \ref{model6} and  \ref{model10},  \ref{model12} and \ref{model13} are T-dual to each other respectively with $b$- and $c$-stacks of brane swapping. 

\begin{table}
	[htb] \footnotesize
	\renewcommand{\arraystretch}{1.0}
	\caption{D6-brane configurations and intersection numbers of Model \ref{model1}, and its MSSM gauge coupling relation is $g^2_a=\frac{8}{3}g^2_b=\frac{8}{3}g^2_c=\frac{8}{5}(\frac{5}{3}g^2_Y)=\frac{8\sqrt{3}}{3}\pi e^{\phi^4}$.} \label{model1}
	\begin{center}
		\begin{tabular}{|c||c|c||c|c|c|c|c|c|c|c|c|c|}
			\hline
			\rm{Model} \ref{model1} & \multicolumn{12}{c|}{$U(12)\times U(2)_L\times U(2)_R\times USp(2)\times USp(2)$}\\
			\hline \hline \rm{stack} & $N$ & $(n^1,l^1)\times (n^2,l^2)\times
			(n^3,l^3)$ & $n_{\Ysymm}$& $n_{\Yasymm}$ & $b$ & $b'$ & $c$ & $c'$& 1 & 2 & 3 & 4 \\
			\hline
			$a$&  24& $(-1,-1)\times (0,-1)\times (-1,-1)$ & 0 & 0  & 1 & 0 & -1 & 0 & 1 & 0 & -1 & 0\\
			$b$&  4& $(2,1)\times (1,-2)\times (1,-1)$ & -1 & -15  & - & - & 0 & 0 & -2 & 0 & -4 & 0 \\
			$c$&  4& $(1,2)\times (1,2)\times (1,-1)$ & 1 & 15  & - & - & - & - & 4 & 0 & 2 & 0\\
			\hline
			1&   2& $(1,0)\times (1,0)\times (2,0)$& \multicolumn{10}{c|}{$X_A =\frac{1}{6} X_B = X_C =\frac{1}{6}X_D$}\\
			3&   2& $(0,-1)\times (1,0)\times (0,2)$ & \multicolumn{10}{c|}{$\beta^g_1=2, \beta^g_3=2$}\\
			&      &	                                                   & \multicolumn{10}{c|}{$\chi_1=1, \chi_2=\frac{1}{6}, \chi_3=2$}\\               
			
			\hline
		\end{tabular}
	\end{center}
\end{table}

\begin{table}
	[htb] \footnotesize
	\renewcommand{\arraystretch}{1.0}
	\caption{D6-brane configurations and intersection numbers of Model \ref{model2}, and its MSSM gauge coupling relation is $g^2_a=\frac{69}{53}g^2_b=\frac{15}{53}g^2_c=\frac{25}{63}(\frac{5}{3}g^2_Y)=\frac{48\sqrt[4]{6}}{53}\pi e^{\phi^4}$.} \label{model2}
	\begin{center}
		\begin{tabular}{|c||c|c||c|c|c|c|c|c|c|c|c|c|}
			\hline
			\rm{Model} \ref{model2} & \multicolumn{12}{c|}{$U(4)\times U(2)_L\times U(6)_R\times USp(2)\times USp(4)$}\\
			\hline \hline \rm{stack} & $N$ & $(n^1,l^1)\times (n^2,l^2)\times
			(n^3,l^3)$ & $n_{\Ysymm}$& $n_{\Yasymm}$ & $b$ & $b'$ & $c$ & $c'$& 1 & 2 & 3 & 4 \\
			\hline
			$a$&  8& $(-1,2)\times (-1,-1)\times (1,-1)$ & 0 & -8  & 0 & 3 & 0 & -1 & -2 & 0 & 1 & 0\\
			$b$&  4& $(-1,-3)\times (1,2)\times (-1,1)$ & 2 & 22  & - & - & 0 & 0 & 6 & 0 & 2 & 0 \\
			$c$&  12& $(0,1)\times (1,-2)\times (1,-1)$ & 1 & -1  & - & - & - & - & -2 & 0 & 0 & 0\\
			\hline
			1&   2& $(1,0)\times (1,0)\times (2,0)$& \multicolumn{10}{c|}{$X_A =\frac{1}{2} X_B =\frac{1}{18} X_C =\frac{1}{24}X_D$}\\
			3 &  4 & $(0,-1)\times (1,0)\times (0,2)$ & \multicolumn{10}{c|}{$\beta^g_1=6, \beta^g_3=-2$}\\
			&      &	                                                   & \multicolumn{10}{c|}{$\chi_1=\frac{1}{6\sqrt{6}}, \chi_2=\frac{\sqrt{\frac{3}{2}}}{2}, \chi_3=2\sqrt{\frac{2}{3}}$}\\               
			
			\hline
		\end{tabular}
	\end{center}
\end{table}

%next table 
\begin{table}
	[htb] \footnotesize
	\renewcommand{\arraystretch}{1.0}
	\caption{D6-brane configurations and intersection numbers of Model \ref{model11}, and its MSSM gauge coupling relation is $g^2_a=\frac{15}{53}g^2_b=\frac{69}{53}g^2_c=\frac{115}{99}(\frac{5}{3}g^2_Y)=\frac{48\sqrt[4]{6}}{53}\pi e^{\phi^4}$.} \label{model11}
	\begin{center}
		\begin{tabular}{|c||c|c||c|c|c|c|c|c|c|c|c|c|}
			\hline
			\rm{Model} \ref{model11} & \multicolumn{12}{c|}{$U(4)\times U(6)_L\times U(2)_R\times USp(2)\times USp(4)$}\\
			\hline \hline \rm{stack} & $N$ & $(n^1,l^1)\times (n^2,l^2)\times
			(n^3,l^3)$ & $n_{\Ysymm}$& $n_{\Yasymm}$ & $b$ & $b'$ & $c$ & $c'$& 1 & 2 & 3 & 4 \\
			\hline
			$a$&  8& $(1,2)\times (-1,1)\times (-1,-1)$ & 0 & 8  & 0 & 1 & 0 & -3 & 2 & 0 & -1 & 0\\
			$b$&  12& $(0,-1)\times (1,2)\times (1,1)$ & -1 & 1  & - & - & 0 & 0 & 2 & 0 & 0 & 0 \\
			$c$&  4& $(1,-3)\times (-1,2)\times (-1,-1)$ & -2 & -22  & - & - & - & - & -6 & 0 & -2 & 0\\
			\hline
			1&   2& $(1,0)\times (1,0)\times (2,0)$& \multicolumn{10}{c|}{$X_A =\frac{1}{2} X_B =\frac{1}{18} X_C =\frac{1}{24}X_D$}\\
			3 &  4 & $(0,-1)\times (1,0)\times (0,2)$ & \multicolumn{10}{c|}{$\beta^g_1=6, \beta^g_3=-2$}\\
			&      &	                                                   & \multicolumn{10}{c|}{$\chi_1=\frac{1}{6\sqrt{6}}, \chi_2=\frac{\sqrt{\frac{3}{2}}}{2}, \chi_3=2\sqrt{\frac{2}{3}}$}\\               
			
			\hline
		\end{tabular}
	\end{center}
\end{table}

\begin{table}
	[htb] \footnotesize
	\renewcommand{\arraystretch}{1.0}
	\caption{D6-brane configurations and intersection numbers of Model \ref{model3}, and its MSSM gauge coupling relation is $g^2_a=\frac{9}{19}g^2_b=\frac{73}{27}g^2_c=\frac{365}{317}(\frac{5}{3}g^2_Y)=\frac{4\sqrt[4]{77}}{57}\pi e^{\phi^4}$.} \label{model3}
	\begin{center}
		\begin{tabular}{|c||c|c||c|c|c|c|c|c|c|c|c|c|}
			\hline
			\rm{Model} \ref{model3} & \multicolumn{12}{c|}{$U(4)\times U(6)_L\times U(2)_R\times USp(2)\times USp(2)$}\\
			\hline \hline \rm{stack} & $N$ & $(n^1,l^1)\times (n^2,l^2)\times
			(n^3,l^3)$ & $n_{\Ysymm}$& $n_{\Yasymm}$ & $b$ & $b'$ & $c$ & $c'$& 1 & 2 & 3 & 4 \\
			\hline
			$a$&  8& $(1,-1)\times (-1,-1)\times (-1,-1)$ & 0 & 4  & 0 & 1 & 0 & -3 & 1 & -1 & 0 & 0\\
			$b$&  12& $(1,2)\times (0,1)\times (-1,-1)$ & -1 & 1  & - & - & 0 & 0 & 2 & 0 & 0 & 0 \\
			$c$&  4& $(1,-2)\times (1,-2)\times (1,1)$ & -1 & -15  & - & - & - & - & -4 & -2 & 0 & 0\\
			\hline
			1&   2& $(1,0)\times (1,0)\times (2,0)$& \multicolumn{10}{c|}{$X_A =\frac{1}{11} X_B =\frac{1}{2} X_C =\frac{1}{14}X_D$}\\
			2 &  2 & $(1,0)\times (0,-1)\times (0,2)$ & \multicolumn{10}{c|}{$\beta^g_1=2, \beta^g_2=-2$}\\
			&      &	                                                   & \multicolumn{10}{c|}{$\chi_1=\frac{\sqrt{\frac{11}{7}}}{2}, \chi_2=\frac{1}{\sqrt{17}}, \chi_3=2\sqrt{\frac{7}{11}}$}\\               
			
			\hline
		\end{tabular}
	\end{center}
\end{table}

%next model

\begin{table}
	[htb] \footnotesize
	\renewcommand{\arraystretch}{1.0}
	\caption{D6-brane configurations and intersection numbers of Model \ref{model15}, and its MSSM gauge coupling relation is $g^2_a=\frac{73}{57}g^2_b=\frac{9}{19}g^2_c=\frac{3}{5}(\frac{5}{3}g^2_Y)=\frac{4(\sqrt[4]{77})^3}{57}\pi e^{\phi^4}$.} \label{model15}
	\begin{center}
		\begin{tabular}{|c||c|c||c|c|c|c|c|c|c|c|c|c|}
			\hline
			\rm{Model} \ref{model15} & \multicolumn{12}{c|}{$U(4)\times U(2)_L\times U(6)_R\times USp(2)\times USp(2)$}\\
			\hline \hline \rm{stack} & $N$ & $(n^1,l^1)\times (n^2,l^2)\times
			(n^3,l^3)$ & $n_{\Ysymm}$& $n_{\Yasymm}$ & $b$ & $b'$ & $c$ & $c'$& 1 & 2 & 3 & 4 \\
			\hline
			$a$&  8& $(1,1)\times (-1,1)\times (-1,1)$ & 0 & -4  & 0 & 3 & -1 & 0 & -1 & 1 & 0 & 0\\
			$b$&  4& $(1,2)\times (1,2)\times (1,-1)$ & 1 & 15  & - & - & 0 & 0 & 4 & 2 & 0 & 0 \\
			$c$&  12& $(1,2)\times (0,1)\times (-1,-1)$ & -1 & 1  & - & - & - & - & 2 & 0 & 0 & 0\\
			\hline
			1&   2& $(1,0)\times (1,0)\times (2,0)$& \multicolumn{10}{c|}{$X_A =\frac{1}{11} X_B =\frac{1}{2} X_C =\frac{1}{14}X_D$}\\
			2 &  2 & $(1,0)\times (0,-1)\times (0,2)$ & \multicolumn{10}{c|}{$\beta^g_1=2, \beta^g_2=-2$}\\
			&      &	                                                   & \multicolumn{10}{c|}{$\chi_1=\frac{\sqrt{\frac{11}{7}}}{2}, \chi_2=\frac{1}{\sqrt{17}}, \chi_3=2\sqrt{\frac{7}{11}}$}\\               
			
			\hline
		\end{tabular}
	\end{center}
\end{table}

\begin{table}
	[htb] \footnotesize
	\renewcommand{\arraystretch}{1.0}
	\caption{D6-brane configurations and intersection numbers of Model \ref{model4}, and its MSSM gauge coupling relation is $g^2_a=\frac{631}{150}g^2_b=\frac{671}{90}g^2_c=\frac{3355}{1612}(\frac{5}{3}g^2_Y)=\frac{8\sqrt[4]{2}(\sqrt[4]{703})^3}{225}\pi e^{\phi^4}$.} \label{model4}
	\begin{center}
		\begin{tabular}{|c||c|c||c|c|c|c|c|c|c|c|c|c|}
			\hline
			\rm{Model} \ref{model4} & \multicolumn{12}{c|}{$U(12)\times U(2)_L\times U(2)_R\times USp(2)$}\\
			\hline \hline \rm{stack} & $N$ & $(n^1,l^1)\times (n^2,l^2)\times
			(n^3,l^3)$ & $n_{\Ysymm}$& $n_{\Yasymm}$ & $b$ & $b'$ & $c$ & $c'$& 1 & 2 & 3 & 4 \\
			\hline
			$a$&  24& $(0,-1)\times (1,-1)\times (-1,1)$ & 0 & 0  & 1 & 0 & 0 & -1 & 0 & 0 & 0 & 0\\
			$b$&  4& $(-1,3)\times (-1,2)\times (1,1)$ & -2 & -22  & - & - & -8 & 0 & 0 & 0 & -2 & 0 \\
			$c$&  4& $(1,-2)\times (3,2)\times (1,-1)$ & -13 & -35  & - & - & - & - & 0 & 0 & 2& 0\\
			\hline
			3&   2& $(0,-1)\times (1,0)\times (0,2)$& \multicolumn{10}{c|}{$X_A = X_B =\frac{4}{37} X_C =\frac{2}{19}X_D$}\\
			&     &                                                   & \multicolumn{10}{c|}{$\beta^g_3=2, \chi_1=2\sqrt{\frac{2}{703}}, \chi_2=\sqrt{\frac{37}{38}}, \chi_3=2\sqrt{\frac{38}{37}}$}\\
			
			\hline
		\end{tabular}
	\end{center}
\end{table}

%next model

\begin{table}
	[htb] \footnotesize
	\renewcommand{\arraystretch}{1.0}
	\caption{D6-brane configurations and intersection numbers of Model \ref{model5}, and its MSSM gauge coupling relation is $g^2_a= \frac{4}{27}g^2_b= \frac{16}{9}g^2_c=\frac{80}{59}(\frac{5}{3}g^2_Y)=\frac{32(\sqrt[4]{5})^3}{27}\pi e^{\phi^4}$.} \label{model5}
	\begin{center}
		\begin{tabular}{|c||c|c||c|c|c|c|c|c|c|c|c|c|}
			\hline
			\rm{Model} \ref{model5} & \multicolumn{12}{c|}{$U(12)\times U(2)_L\times U(2)_R\times USp(4)\times USp(4)\times USp(4)$}\\
			\hline \hline \rm{stack} & $N$ & $(n^1,l^1)\times (n^2,l^2)\times
			(n^3,l^3)$ & $n_{\Ysymm}$& $n_{\Yasymm}$ & $b$ & $b'$ & $c$ & $c'$& 1 & 2 & 3 & 4 \\
			\hline
			$a$&  24& $(1,1)\times (-1,0)\times (-1,1)$ & 0 & 0  & 0 & 1 & -1 & 0 & 0 & 1 & 0 & -1\\
			$b$&  4& $(-1,0)\times (-4,1)\times (-1,1)$ & -5 & 5  & -  & -  & 4 & 0 & 0  & 0 & 1 & 4 \\
			$c$&  4& $(-1,-2)\times (2,-1)\times (-1,-1)$ & 1 & 15  & - & - & - & - & 0 & 4 & -1 & 2\\
			\hline
			2&   4& $(1,0)\times (0,-1)\times (0,2)$& \multicolumn{10}{c|}{$X_A =5 X_B =\frac{5}{4} X_C =5X_D$}\\
			3 &  4 & $(0,-1)\times (1,0)\times (0,2)$ & \multicolumn{10}{c|}{$\beta^g_2=0, \beta^g_3=-4,\beta^g_4=2$}\\
			4 & 4 &$(0,-1)\times (0,1)\times (2,0)$ & \multicolumn{10}{c|}{$\chi_1=\frac{\sqrt{5}}{2}, \chi_2=2\sqrt{5}, \chi_3=\sqrt{5}$}\\

			\hline
		\end{tabular}
	\end{center}
\end{table}

%next model 
\begin{table}
	[htb] \footnotesize
	\renewcommand{\arraystretch}{1.0}
	\caption{D6-brane configurations and intersection numbers of Model \ref{model9}, and its MSSM gauge coupling relation is $g^2_a= \frac{16}{9}g^2_b= \frac{4}{27}g^2_c=\frac{20}{89}(\frac{5}{3}g^2_Y)=\frac{32(\sqrt[4]{5})^3}{27}\pi e^{\phi^4}$.} \label{model9}
	\begin{center}
		\begin{tabular}{|c||c|c||c|c|c|c|c|c|c|c|c|c|}
			\hline
			\rm{Model} \ref{model9} & \multicolumn{12}{c|}{$U(12)\times U(2)_L\times U(2)_R\times USp(4)\times USp(4)\times USp(4)$}\\
			\hline \hline \rm{stack} & $N$ & $(n^1,l^1)\times (n^2,l^2)\times
			(n^3,l^3)$ & $n_{\Ysymm}$& $n_{\Yasymm}$ & $b$ & $b'$ & $c$ & $c'$& 1 & 2 & 3 & 4 \\
			\hline
			$a$&  24& $(-1,1)\times (1,0)\times (-1,-1)$ & 0 & 0  & 1 & 0 & 0 & -1 & 0 & -1 & 0 & 1\\
			$b$&  4& $(-1,2)\times (2,1)\times (-1,1)$ & -1 & -15  & -  & -  & 4 & 0 & 0  & -4 & 1 & -2 \\
			$c$&  4& $(1,0)\times (4,1)\times (-1,-1)$ & 5 & -5  & - & - & - & - & 0 & 0 & -1 & -4\\
			\hline
			2&   4& $(1,0)\times (0,-1)\times (0,2)$& \multicolumn{10}{c|}{$X_A =5 X_B =\frac{5}{4} X_C =5X_D$}\\
			3 &  4 & $(0,-1)\times (1,0)\times (0,2)$ & \multicolumn{10}{c|}{$\beta^g_2=0, \beta^g_3=-4,\beta^g_4=2$}\\
			4 & 4 &$(0,-1)\times (0,1)\times (2,0)$ & \multicolumn{10}{c|}{$\chi_1=\frac{\sqrt{5}}{2}, \chi_2=2\sqrt{5}, \chi_3=\sqrt{5}$}\\    
			\hline
		\end{tabular}
	\end{center}
\end{table}

\begin{table}
	[htb] \footnotesize
	\renewcommand{\arraystretch}{1.0}
	\caption{D6-brane configurations and intersection numbers of Model \ref{model6}, and its MSSM gauge coupling relation is $g^2_a= \frac{49}{15}g^2_b= \frac{4}{75}g^2_c=\frac{20}{233}(\frac{5}{3}g^2_Y)=\frac{32(\sqrt[4]{\frac{13}{3}})^3}{25}\pi e^{\phi^4}$.} \label{model6}
	\begin{center}
		\begin{tabular}{|c||c|c||c|c|c|c|c|c|c|c|c|c|}
			\hline
			\rm{Model} \ref{model6} & \multicolumn{12}{c|}{$U(12)\times U(2)_L\times U(2)_R\times USp(2)\times USp(2)\times USp(2)$}\\
			\hline \hline \rm{stack} & $N$ & $(n^1,l^1)\times (n^2,l^2)\times
			(n^3,l^3)$ & $n_{\Ysymm}$& $n_{\Yasymm}$ & $b$ & $b'$ & $c$ & $c'$& 1 & 2 & 3 & 4 \\
			\hline
			$a$&  24& $(-1,-1)\times (0,1)\times (1,1)$ & 0 & 0  & 0 & 1 & 0 & -1 & 1 & 0 & -1 & 0\\
			$b$&  4& $(-3,2)\times (1,1)\times (-1,-1)$ & 6 & 18 & - & - & 0 &  -9 & 2 & -2 & 3 & 0 \\
			$c$&  4& $(0,-1)\times (1,-4)\times (1,1)$ & 5 &   -5  & - & - & - & - &   -4 & -1& 0 & 0\\
			\hline
			1&   2& $(1,0)\times (1,0)\times (2,0)$& \multicolumn{10}{c|}{$X_A =\frac{1}{4} X_B = X_C =\frac{3}{13}X_D$}\\
			2 &  2 & $(1,0)\times (0,-1)\times (0,2)$ & \multicolumn{10}{c|}{$\beta^g_1=2, \beta^g_2=-3,\beta^g_3=-1$}\\
			3 & 2     &	$(0,-1)\times (1,0)\times (0,2)$ & \multicolumn{10}{c|}{$\chi_1=2\sqrt{\frac{3}{13}}, \chi_2=\frac{\sqrt{\frac{3}{13}}}{2}, \chi_3=\sqrt{\frac{13}{3}}$}\\                   
			
			\hline
		\end{tabular}
	\end{center}
\end{table}

\begin{table}
	[htb] \footnotesize
	\renewcommand{\arraystretch}{1.0}
	\caption{D6-brane configurations and intersection numbers of Model \ref{model10}, and its MSSM gauge coupling relation is $g^2_a= \frac{4}{75}g^2_b= \frac{49}{15}g^2_c=\frac{245}{143}(\frac{5}{3}g^2_Y)=\frac{32(\sqrt[4]{\frac{13}{3}})^3}{25}\pi e^{\phi^4}$.} \label{model10}
	\begin{center}
		\begin{tabular}{|c||c|c||c|c|c|c|c|c|c|c|c|c|}
			\hline
			\rm{Model} \ref{model10} & \multicolumn{12}{c|}{$U(12)\times U(2)_L\times U(2)_R\times USp(2)\times USp(2)\times USp(2)$}\\
			\hline \hline \rm{stack} & $N$ & $(n^1,l^1)\times (n^2,l^2)\times
			(n^3,l^3)$ & $n_{\Ysymm}$& $n_{\Yasymm}$ & $b$ & $b'$ & $c$ & $c'$& 1 & 2 & 3 & 4 \\
			\hline
			$a$&  24& $(-1,-1)\times (0,-1)\times (-1,-1)$ & 0 & 0  & 1 & 0 & -1 & 0 & 1 & 0 & -1 & 0\\
			$b$&  4& $(-1,0)\times (1,-4)\times (-1,1)$ & 5 & -5 & - & - & 0 &  -9 & 0 & 0 & -4 & -1 \\
			$c$&  4& $(2,3)\times (1,1)\times (1,-1)$ & 6 &   18  & - & - & - & - &   3 & 0& 2 & -2\\
			\hline
			1&   2& $(1,0)\times (1,0)\times (2,0)$& \multicolumn{10}{c|}{$X_A =\frac{3}{13} X_B = X_C =\frac{-1}{3}X_D$}\\
			3 &  2 & $(0,-1)\times (1,0)\times (0,2)$ & \multicolumn{10}{c|}{$\beta^g_1=2, \beta^g_2=-3,\beta^g_3=-1$}\\
			4 & 2     &	$(0,-1)\times (1,0)\times (0,2)$ & \multicolumn{10}{c|}{$\chi_1=\frac{\sqrt{\frac{13}{3}}}{2}, \chi_2=\frac{\sqrt{\frac{3}{13}}}{2}, \chi_3=4\sqrt{\frac{3}{13}}$}\\                   
			
			\hline
		\end{tabular}
	\end{center}
\end{table}

%next model 

\begin{table}
	[htb] \footnotesize
	\renewcommand{\arraystretch}{1.0}
	\caption{D6-brane configurations and intersection numbers of Model \ref{model7}, and its MSSM gauge coupling relation is $g^2_a= \frac{16}{3}g^2_b= \frac{16}{3}g^2_c=\frac{80}{41}(\frac{5}{3}g^2_Y)=8\sqrt{\frac{2}{3}}\pi e^{\phi^4}$.} \label{model7}
	\begin{center}
		\begin{tabular}{|c||c|c||c|c|c|c|c|c|c|c|c|c|}
			\hline
			\rm{Model} \ref{model7}I & \multicolumn{12}{c|}{$U(12)\times U(2)_L\times U(2)_R\times USp(2)\times USp(2)\times USp(4)\times USp(4)$}\\
			\hline \hline \rm{stack} & $N$ & $(n^1,l^1)\times (n^2,l^2)\times
			(n^3,l^3)$ & $n_{\Ysymm}$& $n_{\Yasymm}$ & $b$ & $b'$ & $c$ & $c'$& 1 & 2 & 3 & 4 \\
			\hline
			$a$&  24& $(0,-1)\times (-1,1)\times (1,-1)$ & 0 & 0  & 0 & 1 & 0 & -1 & -1 & 1 & 0 & 0\\
			$b$&  4& $(1,4)\times (1,2)\times (1,-1)$ & 3    & 29 & - & -   & 0 & 0 & 8 & 4 & 2 & -1 \\
			$c$&  4& $(-1,4)\times (2,1)\times (-1,1)$ & -3 & -29  & - & - & - & - & -4 & -8 & 1 & -2\\
			\hline
			1&   2& $(1,0)\times (1,0)\times (2,0)$& \multicolumn{10}{c|}{$X_A = X_B =\frac{1}{12} X_C =\frac{1}{12}X_D$}\\
			2 &  2 & $(1,0)\times (0,-1)\times (0,2)$ & \multicolumn{10}{c|}{$\beta^g_1=8, \beta^g_2=8,\beta^g_3=-3, \beta^g_4=-3$}\\
			3 & 4     &	$(0,-1)\times (1,0)\times (0,2)$ & \multicolumn{10}{c|}{$\chi_1=\frac{1}{12}, \chi_2=1, \chi_3=2$}\\    
			4 & 4     &	$(0,-1)\times (0,1)\times (2,0)$ & \multicolumn{10}{c|}{}\\               
			
			\hline
		\end{tabular}
	\end{center}
\end{table}

%next table

\begin{table}
	[htb] \footnotesize
	\renewcommand{\arraystretch}{1.0}
	\caption{D6-brane configurations and intersection numbers of Model \ref{model8}, and its MSSM gauge coupling relation is $g^2_a=\frac{221}{45}g^2_b=\frac{245}{27}g^2_c=\frac{1225}{571}(\frac{5}{3}g^2_Y)=\frac{8(\sqrt[4]{506})^3}{135}\pi e^{\phi^4}$.}\label{model8}
	\begin{center}
		\begin{tabular}{|c||c|c||c|c|c|c|c|c|c|c|c|c|}
			\hline
			\rm{Model} \ref{model8} & \multicolumn{12}{c|}{$U(12)\times U(2)_L\times U(2)_R\times USp(4)$}\\
			\hline \hline \rm{stack} & $N$ & $(n^1,l^1)\times (n^2,l^2)\times
			(n^3,l^3)$ & $n_{\Ysymm}$& $n_{\Yasymm}$ & $b$ & $b'$ & $c$ & $c'$& 1 & 2 & 3 & 4 \\
			\hline
			$a$&  24& $(-1,0)\times (1,1)\times (-1,1)$ & 0 & 0         & 1 & 0 & -1& 0 & 0 & 0 & 0 & 0\\
			$b$&  4& $(-4,-1)\times (-2,-1)\times (-1,-1)$& -3 & -29  & - & - & 0 & -16 & 1 & 0 & 0 & 0 \\
			$c$&  4& $(2,-1)\times (-2,-3)\times (-1,-1)$ & 13 & 35  & - & - & - & - & 3       & 0 & 0 & 0\\
			\hline
			1&   4& $(1,0)\times (1,0)\times (2,0)$& \multicolumn{10}{c|}{$X_A =\frac{22}{23} X_B =11X_C =11X_D$}\\
			&   &  & \multicolumn{10}{c|}{$\beta^g_1=-2, \chi_1=\sqrt{\frac{253}{2}}, \chi_2=\sqrt{\frac{22}{23}}, \chi_3=2\sqrt{\frac{22}{23}}$}\\
			
			\hline
		\end{tabular}
	\end{center}
\end{table}

%next model

\begin{table}
	[htb] \footnotesize
	\renewcommand{\arraystretch}{1.0}
	\caption{D6-brane configurations and intersection numbers of Model \ref{model12}, and its MSSM gauge coupling relation is $g^2_a=\frac{3}{2}g^2_b=\frac{7}{6}g^2_c=\frac{35}{32}(\frac{5}{3}g^2_Y)=\frac{8\sqrt[4]{2}}{3}\pi e^{\phi^4}$.} \label{model12}
	\begin{center}
		\begin{tabular}{|c||c|c||c|c|c|c|c|c|c|c|c|c|}
			\hline
			\rm{Model} \ref{model12} & \multicolumn{12}{c|}{$U(4)\times U(6)_L\times U(2)_R\times USp(2)$}\\
			\hline \hline \rm{stack} & $N$ & $(n^1,l^1)\times (n^2,l^2)\times
			(n^3,l^3)$ & $n_{\Ysymm}$& $n_{\Yasymm}$ & $b$ & $b'$ & $c$ & $c'$& 1 & 2 & 3 & 4 \\
			\hline
			$a$&  8& $(0,-1)\times (1,-1)\times (-1,1)$ & 0 & 0  & 0 & 1 & 0 & -3 & 0 & 0 & 0 & 0\\
			$b$&  12& $(1,0)\times (1,2)\times (1,-1)$ & -1 & 1  & - & - & 0 & 0 & 0 & 0 & 2 & 0 \\
			$c$&  4& $(-1,-1)\times (-1,2)\times (1,-1)$ & 0 & -8  & - & - & - & - & 0 & 0 & -2& 0\\
			\hline
			3&   2& $(0,-1)\times (1,0)\times (0,2)$& \multicolumn{10}{c|}{$X_A = X_B =4 X_C =2 X_D$}\\
			&     &                                                   & \multicolumn{10}{c|}{$\beta^g_3=-2, \chi_1=2\sqrt{2}, \chi_2=\frac{1}{\sqrt{2}}, \chi_3=2\sqrt{2}$}\\
			
			\hline
		\end{tabular}
	\end{center}
\end{table}

\begin{table}
	[htb] \footnotesize
	\renewcommand{\arraystretch}{1.0}
	\caption{D6-brane configurations and intersection numbers of Model \ref{model13}, and its MSSM gauge coupling relation is $g^2_a=\frac{7}{6}g^2_b=\frac{3}{2}g^2_c=\frac{5}{4}(\frac{5}{3}g^2_Y)=\frac{8\sqrt[4]{2}}{3}\pi e^{\phi^4}$.} \label{model13}
	\begin{center}
		\begin{tabular}{|c||c|c||c|c|c|c|c|c|c|c|c|c|}
			\hline
			\rm{Model} \ref{model13} & \multicolumn{12}{c|}{$U(4)\times U(2)_L\times U(6)_R\times USp(2)$}\\
			\hline \hline \rm{stack} & $N$ & $(n^1,l^1)\times (n^2,l^2)\times
			(n^3,l^3)$ & $n_{\Ysymm}$& $n_{\Yasymm}$ & $b$ & $b'$ & $c$ & $c'$& 1 & 2 & 3 & 4 \\
			\hline
			$a$&  8& $(1,-1)\times (-1,0)\times (-1,-1)$ & 0 & 0  & 3 & 0 & 0 & -1 & 0 & 0 & 0 & 0\\
			$b$&  4& $(-2,-1)\times (1,-1)\times (-1,1)$ & 0 & -8  & - & - & 0 & 0 & 0 & 0 & -2 & 0 \\
			$c$&  12& $(-2,-1)\times (0,1)\times (1,1)$ & 1 & -1  & - & - & - & - & 0 & 0 & -2& 0\\
			\hline
			3&   2& $(0,-1)\times (1,0)\times (0,2)$& \multicolumn{10}{c|}{$X_A = \frac{1}{2}X_B =2 X_C =\frac{1}{2} X_D$}\\
			&     &                                                   & \multicolumn{10}{c|}{$\beta^g_3=-2, \chi_1=\sqrt{2}, \chi_2=\frac{1}{2\sqrt{2}}, \chi_3=2\sqrt{2}$}\\
			
			\hline
		\end{tabular}
	\end{center}
\end{table}

\begin{table}
	[htb] \footnotesize
	\renewcommand{\arraystretch}{1.0}
	\caption{D6-brane configurations and intersection numbers of Model \ref{model14}, and its MSSM gauge coupling relation is $g^2_a=\frac{53}{9}g^2_b=\frac{5}{63}g^2_c=\frac{25}{199}(\frac{5}{3}g^2_Y)=\frac{8(\sqrt[4]{110})^3}{63}\pi e^{\phi^4}$.}\label{model14}
	\begin{center}
		\begin{tabular}{|c||c|c||c|c|c|c|c|c|c|c|c|c|}
			\hline
			\rm{Model} \ref{model14} & \multicolumn{12}{c|}{$U(12)\times U(2)_L\times U(2)_R\times USp(4)$}\\
			\hline \hline \rm{stack} & $N$ & $(n^1,l^1)\times (n^2,l^2)\times
			(n^3,l^3)$ & $n_{\Ysymm}$& $n_{\Yasymm}$ & $b$ & $b'$ & $c$ & $c'$& 1 & 2 & 3 & 4 \\
			\hline
			$a$&  24& $(1,1)\times (1,0)\times (1,-1)$ & 0 & 0         & 0 & 1 & 0& -1 & 0 & 0 & 0 & 0\\
			$b$&  4& $(-4,3)\times (1,-1)\times (1,-1)$& 16 & 32  & - & - & 0 & -16 & -3 & 0 & 0 & 0 \\
			$c$&  4& $(0,-1)\times (5,1)\times (1,-1)$ & 6 & -6  & - & - & - & - & -1      & 0 & 0 & 0\\
			\hline
			1&   4& $(1,0)\times (1,0)\times (2,0)$& \multicolumn{10}{c|}{$X_A =5 X_B =\frac{10}{11}X_C =5X_D$}\\
			&   &  & \multicolumn{10}{c|}{$\beta^g_1=-2, \chi_1=\sqrt{\frac{10}{11}}, \chi_2=\sqrt{\frac{55}{2}}, \chi_3=2\sqrt{\frac{10}{11}}$}\\
			
			\hline
		\end{tabular}
	\end{center}
\end{table}

\end{appendix}

\end{document}